\documentclass[a4paper,11pt]{article}
\pdfoutput=1
\usepackage{graphicx,rotating,hyperref,slashed,amsmath,xcolor,amssymb,amsfonts,colortbl,cite, subfigure,float}
\usepackage{multirow}
\makeatletter 
\usepackage{graphics}
\usepackage{graphicx} 
\graphicspath{{plots/}}
\usepackage{comment}
\usepackage{url}

\hypersetup{colorlinks,bookmarksopen,bookmarksnumbered,
linkcolor=blus,pdfstartview=FitH,urlcolor=rossos,citecolor=verde}
\allowdisplaybreaks

\def\lsim{\mathrel{\rlap{\lower3pt\hbox{\hskip0pt$\sim$}}
   \raise1pt\hbox{$<$}}}         
\def\gsim{\mathrel{\rlap{\lower4pt\hbox{\hskip1pt$\sim$}}
   \raise1pt\hbox{$>$}}}         

 \newcommand{\sfootnote}[1]{} 
\definecolor{bluc}{cmyk}{1,1,0,0.1}
\definecolor{rossoCP3}{cmyk}{0,.88,.77,.40}
\definecolor{rosso}{cmyk}{0,1,1,0.4}
\definecolor{rossos}{cmyk}{0,1,1,0.55}
\definecolor{rossoc}{cmyk}{0,1,1,0.2}
\definecolor{verdes}{cmyk}{0.92,0,0.59,0.4}

\hypersetup{colorlinks, bookmarksopen, bookmarksnumbered,
citecolor=verdes, linkcolor=bluc, pdfstartview=FitH, urlcolor=rossos}

\newcommand{\mio}[1]{}

\definecolor{Gray}{gray}{0.95}

\def\bx{{\bf x}}

\def\bn{{\bf n}}
\def\bv{{\bf v}}

\usepackage{multicol}
\usepackage{color}
\definecolor{rosso}{cmyk}{0,1,1,0.4}
\definecolor{rossos}{cmyk}{0,1,1,0.55}
\definecolor{rossoc}{cmyk}{0,1,1,0.2}
\definecolor{blu}{cmyk}{1,1,0,0.3}
\definecolor{blus}{cmyk}{1,1,0,0.6}
\definecolor{bluc}{cmyk}{1,1,0,0.1}
\definecolor{verde}{cmyk}{0.92,0,0.59,0.25}
\definecolor{verdec}{cmyk}{0.92,0,0.59,0.15}
\definecolor{verdes}{cmyk}{0.92,0,0.59,0.4}

\def\circa#1{\,\raise.3ex\hbox{$#1$\kern-.75em\lower1ex\hbox{$\sim$}}\,}

\newcommand{\beq}{\begin{equation}}
\newcommand{\eeq}{\end{equation}}

\newcommand{\bea}{\begin{eqnarray}}
\newcommand{\eea}{\end{eqnarray}}
\newcommand{\be}{\begin{equation}}
\newcommand{\ee}{\end{equation}}
\newfam\rsfsfam
\def\mathscr#1{{\fam\rsfsfam\relax#1}}

\def\circa#1{\,\raise.3ex\hbox{$#1$\kern-.75em\lower1ex\hbox{$\sim$}}\,}
\makeatletter

\def\hhref#1{\href{http://arxiv.org/abs/#1}{arXiv:#1}} 

\newcommand{\doi}[1]{\href{http://dx.doi.org/#1}{[doi]}}

\def\hhref#1{\href{http://arxiv.org/abs/#1}{arXiv:#1}} 
 
\def\art{\@ifnextchar[{\eart}{\oart}}
\def\eart[#1]#2#3#4#5#6{{\rm #2}, {\em #3 \bf #4} {\rm (#6) #5} ({\em #1})}

\def\article{\@ifnextchar[{\earticle}{\oarticle}}
\def\oarticle#1#2#3#4#5#6{{\rm #1}, {\em ``#6''}, {\rm #2 #3 (#5) #4}}
\def\earticle[#1]#2#3#4#5#6#7{{\rm #2}, {\em ``#7''}, {\rm #3 #4 (#6) #5}  [\hhref{#1}]}
\def\hepart[#1]#2{{\rm #2, \em#1}}
\def\heparticle[#1]#2#3{#2, {\em ``#3''} [\hhref{#1}]}

\newcounter{alphaequation}[equation]
\def\thealphaequation{\theequation\hbox to
0.6em{\hfil\alph{alphaequation}\hfil}}
\def\eqnsystem#1{
\def\@eqnnum{{\rm (\thealphaequation)}}
\def\@@eqncr{\let\@tempa\relax \ifcase\@eqcnt \def\@tempa{& & &} \or
  \def\@tempa{& &}\or \def\@tempa{&}\fi\@tempa
  \if@eqnsw\@eqnnum\refstepcounter{alphaequation}\fi
\global\@eqnswtrue\global\@eqcnt=0\cr}
\refstepcounter{equation} \let\@currentlabel\theequation \def\@tempb{#1}
\ifx\@tempb\empty\else\label{#1}\fi
\refstepcounter{alphaequation}
\let\@currentlabel\thealphaequation
\global\@eqnswtrue\global\@eqcnt=0 \tabskip\@centering\let\\=\@eqncr
$$\halign to \displaywidth\bgroup \@eqnsel\hskip\@centering
$\displaystyle\tabskip\z@{##}$&\global\@eqcnt\@ne
\hskip2\arraycolsep\hfil${##}$\hfil& \global\@eqcnt\tw@\hskip2\arraycolsep
$\displaystyle\tabskip\z@{##}$\hfil
\tabskip\@centering&\llap{##}\tabskip\z@\cr}
\def\endeqnsystem{\@@eqncr\egroup$$\global\@ignoretrue} \makeatother

\definecolor{fiorentina}{rgb}{.5,0,.5}

\definecolor{darkgreen}{RGB}{0, 100, 0}
\definecolor{cred}{RGB}{180,50,40} 
   
\begin{document}

\vspace{1truecm}
 \begin{center}
\boldmath

\begin{flushright}
 DESY 19-161
\end{flushright}
\vspace{5mm}

{\textbf{\Large 
Measuring the net circular polarization of the \\
\smallskip
stochastic gravitational wave background  with interferometers}}
\unboldmath

\unboldmath

\bigskip

\vspace{0.1truecm}
\date\today
{\bf Valerie Domcke\,$^a$, Juan Garc\'ia-Bellido\,$^{b}$,  Marco Peloso\,$^{c,d}$,  Mauro Pieroni\,$^{b,e}$,\\  Angelo Ricciardone\,$^{c}$, Lorenzo Sorbo\,$^f$, Gianmassimo Tasinato\,$^g$}
 \\[8mm]
 {\it $^a$ Deutsches Elektronen Synchrotron (DESY), 22607 Hamburg, Germany}\\[1mm]
{\it $^b$ Instituto de F\'{\i}sica Te\'orica UAM/CSIC, Nicol\'as Cabrera 13, Universidad
Aut\'onoma de Madrid, Cantoblanco 28049 Madrid, Spain}\\[1mm]
{\it $^c$ INFN, Sezione di Padova, via Marzolo 8, I-35131, Padova, Italy}  \\[1mm]
 {\it $^d$ Dipartimento di Fisica e Astronomia ``G. Galilei'', Universit\`a degli Studi di Padova,\\ via Marzolo 8, I-35131, Padova, Italy}  \\[1mm]
{\it $^e$ Theoretical Physics, Blackett Laboratory, Imperial College, London, SW7 2AZ,\\ United Kingdom}\\[1mm]
{\it $^f$ Amherst Center for Fundamental Interactions, Department of Physics, University of Massachusetts, Amherst, MA 01003, USA}\\[1mm]
{\it $^g$  Department of Physics, Swansea University, Swansea, SA2 8PP, United Kingdom
}\\[1mm]
\vspace{1cm}

\end{center}

\begin{abstract}
Parity violating interactions in the early Universe can source a stochastic gravitational wave background (SGWB) with a net circular polarization.
In this paper,  we study possible ways to search for circular polarization  of 
the SGWB with interferometers.
 Planar detectors are unable to measure the net circular polarization of an isotropic SGWB.  We discuss the possibility of using
the dipolar anisotropy kinematically induced by the motion of the solar system  with respect  to the cosmic reference frame to measure the net circular polarization of the SGWB  with planar detectors. We apply this approach to LISA, re-assessing previous analyses by means of a more detailed computation and using the most recent  instrument specifications, and to the Einstein Telescope (ET), estimating for the first time its sensitivity  to circular polarization. We find that both LISA and ET, despite operating at different frequencies, could detect net circular polarization with a signal-to-noise ratio of order one in a SGWB with amplitude $h^2 \Omega_\text{GW} \simeq 10^{-11}$. We also investigate the case of a network of ground based detectors. We present fully analytical, covariant formulas for the detector overlap functions in the presence   of circular polarization. Our formulas   do not rely on particular   choices of reference frame, and can be applied to  interferometers with arbitrary angles among their arms. 
    \end{abstract}

\newpage
    \section{Introduction}

A direct detection of the SGWB  represents a major future target of gravitational wave (GW) experiments working at interferometer scales.   The  characterization of the SGWB  properties, and the corresponding
  detection strategies,  are   essential for   distinguishing between a   cosmological and  an astrophysical origin of the signal. See e.g.\  \cite{Allen:1996vm,Caprini:2018mtu,Bartolo:2016ami,Regimbau:2011rp,Maggiore:2018sht,Romano:2016dpx} for comprehensive reviews on theoretical and experimental aspects of the physics of SGWBs. Among the  properties that can characterize a SGWB is an intrinsic circular polarization, associated with an asymmetry in the amplitude of GWs of left and right polarizations. 

The astrophysical SGWB is a  combination of several independent signals from uncorrelated sources. Therefore,
 we do not expect the astrophysical SGWB to carry a net polarization.    
On the other hand, cosmological  SGWBs can be produced coherently (for example, the SGWB from inflation): if this coherence  is accompanied by interactions that violate parity, then a cosmological  SGWB with net circular polarization can be generated.   In fact,  a sizable degree of polarization can be generated  in well-motivated models of inflation with spontaneous parity violation, manifesting itself e.g.\ in Chern-Simons  couplings between  the inflaton $\phi$ and  curvature (as $\phi\, R\,\tilde R$, \cite{Lue:1998mq,Jackiw:2003pm,Contaldi:2008yz,Bartolo:2017szm}) or gauge fields (as $\phi\,F\,\tilde F$, see e.g.~\cite{Sorbo:2011rz,Barnaby:2011qe,Adshead:2013nka, Shiraishi:2013kxa,Namba:2015gja,Dimastrogiovanni:2016fuu}).  Such a scenario, and its consequences for CMB polarization experiments, is the subject of active research, see e.g.~\cite{Alexander:2009tp,Maleknejad:2012fw,Pajer:2013fsa,Shiraishi:2016ads} for reviews.  Interestingly, recent  numerical  analysis \cite{Pol:2019yex}  show that  post-inflationary physics associated with  magnetohydrodynamic turbulence, in the presence of helical initial magnetic fields,  can also give rise to net circular polarization  of a SGWB potentially detectable with LISA. In this work, we will study the prospects for detecting a net circular polarization in the SGWB in GW interferometry experiments. A positive detection would provide a smoking gun for parity violating effects and for a cosmological origin of the SGWB signal.
 
 \smallskip
 
It has been  proven \cite{Seto:2007tn,Seto:2008sr,Smith:2016jqs} that parity violating effects in an isotropic SGWB can {\it not} be  detected by correlating  a system of coplanar detectors. A planar interferometer responds in the same way to a left-handed GW of wave vector $\vec{k}$ and to a right-handed GW of the same amplitude and of wave vector $\vec{k}_p$, obtained from $\vec{k}$ by changing sign of the component of $\vec{k}$ perpendicular to the plane of the detector. In particular, this is the case for LISA and ET, which are planar instruments.  A way out of this argument is provided by an anisotropic SGWB \cite{Seto:2006hf,Seto:2006dz}, since in this case the GW arriving from the direction  $\vec{k}$ have a different amplitude than those from the  $\vec{k}_p$ direction. Moreover, this problem is not present when one correlates signals from different GW detectors which do not lie on the same plane \cite{Seto:2007tn,Seto:2008sr,Crowder:2012ik}, as is the case for a network of  ground-based interferometers.

In this work, we start from the consideration that a SGWB that is (statistically) isotropic in one frame ${\cal O}$ is not (statistically) isotropic in any other frame that is boosted with respect to ${\cal O}$. This is true for any stochastic background, and this is for example the origin of the CMB dipole, which is induced kinematically by the motion of the solar system frame with respect to the cosmic reference frame. The latter is defined to be the one in which the CMB is statistically isotropic, and it is the rest frame of the cosmic fluid. It is reasonable to assume that this is also the frame in which the SGWB is isotropic.~\footnote{In fact, the analysis described in this paper will allow us to test this hypothesis, if the SGWB has a net circular polarization.} The fact  that  measurements of  parity odd SGWB anisotropies allows the detection of circular polarization  was already noticed and developed in \cite{Seto:2006hf,Seto:2006dz}. In the present work, we present a more detailed computation for the LISA instrument, discussing in full extent the properties of the instrument response functions under parity symmetry in the presence of a dipolar anisotropy, and clarifying the relation between these properties and the SGWB circular polarization. Using the most up-to-date LISA instrument specifications, and taking into account the full frequency band of the instrument, we re-assess  the evaluation of the magnitude of the signal-to-noise ratio associated with measurements of the SGWB circular polarization, obtaining a result about one order of magnitude greater that that of  \cite{Seto:2006hf}.

This analysis can be readily extended to the the proposed ground-based Einstein Telescope (ET).~\footnote{ET will be a ground-based interferometer with a triangular shape, like LISA, with the difference that the arm length is $L=10\,$km. It  will be an observatory of the third generation aiming to reach a sensitivity for GW signals emitted by astrophysical and cosmological sources about a factor of 10 better than the advanced detectors currently operating. It will be formed by three detectors, each in turn composed of two interferometers (xylophone configuration)~\cite{Punturo:2010zz, Sathyaprakash:2012jk}. }  A single third-generation telescope of this type features a planar configuration similar to that of LISA. Using also in this case the kinematically induced dipole, we estimate 
for the first time the signal-to-noise ratio (SNR) for this measurement at ET. We find that both LISA and ET, despite operating at different frequencies, could detect net circular polarization with a signal-to-noise ratio of order one in a SGWB with amplitude $h^2 \Omega_\text{GW} \simeq 10^{-11}$. 

We then consider correlations of ground-based interferometers. In this case, as mentioned above, a net circular polarization can already be measured from the SGWB monopole (namely, from its statistically isotropic component), since a network of two or more detectors is generically not coplanar. Such an analysis was already performed in \cite{Crowder:2012ik} for the second and third generation ground-based interferometers.~\footnote{To detect chirality, we need to measure $P_L$ and $P_R$ separately, so at least three interferometers are needed. Two interferometers are enough if one assumes as an input the spectral form of the signal, as in this case the measurements at different frequencies can be combined together \cite{Crowder:2012ik}. For planar interferometers such as LISA and ET, the needed plurality of measurements is guaranteed by the different time-delay-interferometers at their vertices.} While in  \cite{Crowder:2012ik} a numerical evaluation of the parity-dependent overlap functions was employed, in this paper we compute, for the first time, the full analytic form of these functions (the overlap functions for parity even backgrounds were computed analytically in~\cite{Allen:1997ad}). We present  `covariant' analytic  formulas for overlap functions describing correlations among ground based interferometers in the small antenna limit (which applies to all existing ground-based interferometers), also including the kinematically induced dipolar anisotropy. Our expressions are valid for any amount of polarization of the SGWB (namely, we provide separate formulas for the left-handed and the right-handed GW), they do not rely on any special choices of frame (this is why we call them covariant), and they hold for arbitrary detector shape (namely, they are not limited to interferometers with orthogonal arms). While the angular integrals necessary to obtain the overlap functions can be also computed numerically \cite{Crowder:2012ik}, evaluating the analytic formulae given here is significantly faster, and we hope that it might speed up such analyses. 

The structure of the paper is the following: in Section~\ref{sec: boostpowerspectrum} we  compute the GW two-point function for a detector which is boosted with respect to a frame in which the SGWB is isotropic; in Section~\ref{sec: sec_LISA} we present the dipole response functions for measuring the net circular polarization of the SGWB with LISA. We turn to ground-based detectors in Section~\ref{sec:ground_based}, considering both the case of cross-correlations among a network of (not coplanar) interferometers, for which already the monopole overlap function is sensitive to chirality, as well as the proposed Einstein Telescope, which can measure chirality upon taking into account the kinematic dipole. We  conclude in Section~\ref{sec: conclusions}. Four appendices provide further technical details. In App.~\ref{app:GW-polarization-operators}, we specify the GW polarization operators employed in this work. App.~\ref{app:comparison} compares our findings with those of ref. \cite{Seto:2006hf} for the measurement of the SGWB circular polarization with LISA. App.~\ref{app:detectors} lists the position of the ground-based detectors considered, and App.~\ref{app:ground-analytic} contains the derivation of the analytical expressions for the monopole and dipole overlap functions for ground-based detectors.

 \section{Dipolar anisotropy of a cosmological  SGWB }
\label{sec: boostpowerspectrum}

Let us assume that there exists a frame in which the  SGWB is (statistically) isotropic. It is natural to associate this frame to the cosmological frame, in which the CMB is isotropic. The peculiar motion of the solar system in this frame will kinematically make the observed SGWB anisotropic, as is this the case for the CMB, where it is found that our local system is moving with speed $v\,=\, 1.23\times10^{-3}$ in a direction $( \phi_E,\,\theta_E)\,=\,(172^{\circ},\,-11^{\circ})$ in ecliptic coordinates (see e.g. \cite{Tanabashi:2018oca}). The possibility to detect a kinematically-induced dipolar anisotropy with ground based experiments was first quantitatively explored in \cite{Allen:1996gp}, and more recently re-assessed in  \cite{Kudoh:2005as} for the space-based experiment DECIGO. In this Section, we derive general formulas describing how a dipolar anisotropy is induced on an otherwise isotropic SGWB. In Section \ref{sec: sec_LISA}, we  use these results to study how such dipolar anisotropy can enable the detection of the net circular polarization of a SGWB with the LISA instrument.

\smallskip

We  compute the GW two-point function seen by an observer who is moving with a constant velocity $\vec{v}$ with respect to a frame in which the SGWB is isotropic. The motion with velocity $\vec{v}$ of the observer generates a dipole in the observed GW power spectrum at order $v$, a quadrupole at order $v^2$ and so on. Under the assumption that $v\ll 1$ (as it is  the case if the isotropic frame of the SGWB and of the CMB coincide), we  only focus on the dipole component, considering  terms up to $\mathcal{O}(v)$.

We start the computation by considering a frame $\left\{ t ,\, \vec{x} \right\}$ in which the SGWB is isotropic. In this frame, we decompose the tensor field into modes of definite circular polarization, with $\lambda = \pm 1$ denoting right- and left-handed modes, respectively,
\begin{align}\label{eq:h_deco}
h_{ij}( t ,\, \vec{x})=\int {d^3k} \, e^{-2\pi i\vec{k}\cdot\vec{x}}\sum_\lambda e_{ij,\lambda}(\hat{k})\,h^\lambda(t,\vec{k})\,,
\end{align}
where the GW polarization operators in the chiral basis $e_{ab,\lambda}(  {\hat k})$ are introduced in Appendix \ref{app:GW-polarization-operators}. The mode momentum-space operators of definite helicity satisfy the condition $h^\lambda(t,\vec{k})=h^\lambda(t,-\vec{k})^*$ which, together with the property (\ref{e-properties}), ensures that the expression (\ref{eq:h_deco}) is real. This expression satisfies the wave equation for a massless particle, which is solved by 
\begin{align}
h^\lambda(t,\vec{k})=A^\lambda_{\vec{k}}\,\cos(2\pi k\,t)+B^\lambda_{\vec{k}}\,\sin(2\pi k\,t) \;, 
\end{align}
where  $A^\lambda_{\vec{k}}=(A^\lambda_{-\vec{k}})^*$ and $B^\lambda_{\vec{k}}=(B^\lambda_{-\vec{k}})^*$ are stochastic variables that obey
\begin{align}\label{eq:abcorrelators}
\langle A^\lambda_{\vec{k}}\,A^{\lambda'}_{\vec{k}'}\rangle=\langle B^{\lambda}_{\vec{k}}\,B^{\lambda'}_{\vec{k}'}\rangle=\frac{P_\lambda(k)}{4\pi k^3}\delta_{\lambda\lambda'}\delta(\vec{k}+\vec{k}')\,,\qquad \langle A^{\lambda}_{\vec{k}}\,B^{\lambda'}_{\vec{k}'}\rangle=0\,, 
\end{align}
where $P_\lambda(k)$ is the GW helicity-$\lambda$ power spectrum, depending only on the absolute value $k$ due to statistical isotropy. We note that, with our $2 \pi$ convention, $k = \vert \vec{k} \vert$ is the frequency of the mode. Moreover, we have 
\begin{equation}
{\rm no \; net \; circular \; polarization  } \;\;\; \Leftrightarrow \;\;\; P_R \left( k \right) =  P_L \left( k \right) 
 \;\;\; \Leftrightarrow \;\;\; \sum_\lambda \lambda \, P_\lambda \left( k \right) = 0 \,. 
 \end{equation} 
Equations~(\ref{eq:abcorrelators}) derive from the requirement that the equal time correlator takes the time-independent form\footnote{Here we are considering the present-day SGWB, evaluated at times relevant for the detection. When considering cosmological time scales (e.g.\ when comparing with the primordial power spectrum), the expansion of the Universe must be taken into account, encoded in the cosmic transfer function \label{ftn:cosmic}.}
\begin{align}
\langle h^\lambda(t,\vec{k})h^{\lambda'}(t,\vec{k}')\rangle\equiv \frac{P_\lambda(k)}{4\pi k^3}\delta_{\lambda\lambda'}\delta(\vec{k}+\vec{k}') \,.
\label{eq:hh}
\end{align}

The gravitational wave correlator at arbitrary times then reads
\begin{align}\label{eq:unequal_corr}
\langle h_{ij}(\vec{x},\,t)h_{i'j'}(\vec{x}\,',\,t')\rangle=&\sum_{\sigma}\int\frac{d^3k}{4\pi k^3}\,{e^{-2\pi i\vec{k}\cdot(\vec{x}-\vec{x}\,')}}\,e_{ij,\sigma}(\hat{k})e_{i'j',\sigma}(-\hat{k})P^{\sigma}(k)\,\cos(2\pi k(t-t'))\nonumber\\
=&\frac{1}{2}\sum_{\sigma}\int\frac{d^3k}{4\pi k^3}\,e^{-2\pi i\vec{k}\cdot(\vec{x}-\vec{x}\,')+2\pi ik(t-t')}\,\,e_{ij,\sigma}(\hat{k})e_{i'j',\sigma}(-\hat{k})P^{\sigma}(k)\nonumber\\
&+\frac{1}{2}\sum_{\sigma}\int\frac{d^3k}{4\pi k^3}\,{e^{-2\pi i\vec{k}\cdot(\vec{x}-\vec{x}\,')-2\pi ik(t-t')}}\,e_{ij,\sigma}(\hat{k})e_{i'j',\sigma}(-\hat{k})P^{\sigma}(k)\,.
\end{align}

We now perform a boost to a frame $\left\{ \tau ,\, \vec{y} \right\}$ that is moving with constant velocity $\vec{v}$, directed along the first coordinate, with respect to the  $\left\{ t ,\, \vec{x} \right\}$ frame 
\begin{align}\label{eq:boost}
&t=\gamma(\tau-v\,y_1)  \;,\;\; x_1=\gamma(y_1-v\,\tau) \;,\;\; x_2=y_2 \;,\;\; 
x_3=y_3 \,, 
\end{align}
where  $\gamma\equiv 1/\sqrt{1-v^2}$. Being a rank-2 tensor, $h_{ij}$  transforms as 
\begin{eqnarray} 
h_{ij}(x_1,\,x_2,\,x_3,\,t) &=& h_{ab}(\gamma(y_1-v\,\tau),\,y_2,\,y_3,\,\gamma(\tau-v\,y_1))\frac{\partial y_a}{\partial x_i}\frac{\partial y_b}{\partial x_j} \nonumber\\ 
&\simeq& h_{ij}(\gamma(y_1-v\,\tau),\,y_2,\,y_3,\,\gamma(\tau-v\,y_1)) + {\mathcal{O} } \left( v^2 \right) \,.
\end{eqnarray} 
Let us perform this transformation on the decomposition (\ref{eq:unequal_corr}). To preserve the same plane wave structure of the phase in the decomposition, we simultaneously perform a change in the integration variable, which can be also thought of as a boost on the momenta, with opposite signs of the boost parameter depending on whether we are in the negative (second line of eq.~(\ref{eq:unequal_corr}), $\vec k \mapsto \vec q$) or positive (third  line of eq.~(\ref{eq:unequal_corr}), $\vec k \mapsto \vec p$) frequency component of the unequal-time correlator,
\begin{align}
& {\rm second\ line\ of\ eq.~(\ref{eq:unequal_corr})}  & {\rm third\ line\ of\ eq.~(\ref{eq:unequal_corr})}  \nonumber \\
& \left\{
\begin{array}{l}
k_1=\gamma(q_1-v\,q)\\
k_2=q_2\\
k_3=q_3\\
k=\gamma(q-v\,q_1)
\end{array}\right.
\,,
& \left\{
\begin{array}{l}
k_1=\gamma(p_1+v\,p)\\
k_2=p_2\\
k_3=p_3\\
k=\gamma(p+v\,p_1)
\end{array}\right.
\end{align}
with $q\equiv |\vec{q}|$ and $ p\equiv |\vec{p}| $.
Therefore, the unequal time correlator in the boosted frame can be written as 
\begin{align}\label{eq:corr_mixed_variables}
\langle h_{ij}(\vec{y},\,\tau)h_{i'j'}(\vec{y}\,',\,\tau')\rangle=&\frac{1}{2}\sum_\sigma\int\frac{d^3k}{4\pi k^{3}}\,{e^{-2\pi i\vec{q}\cdot(\vec{y}-\vec{y}\,')+2\pi iq(\tau-\tau')}}\,e_{ij,\sigma}(\hat{k})e_{i'j',\sigma}(-\hat{k})P^{\sigma}(k)\nonumber\\
&+\frac{1}{2}\sum_\sigma\int\frac{d^3k}{4\pi k^{3}}\,{e^{-2\pi i\vec{p}\cdot(\vec{y}-\vec{y}\,')-2\pi ip(\tau-\tau')}}e_{ij,\sigma}(\hat{k})e_{i'j',\sigma}(-\hat{k})P^{\sigma}(k)\,,
\end{align}
where the dependence on the velocity $\vec{v}$ is hidden in the relation between the variables $\vec{q}$, $\vec{p}$ and $\vec{k}$.

In the following, we perform explicit computations only on the first term on the right hand side of eq.~(\ref{eq:corr_mixed_variables}), since the second one is obtained from the first one with the replacements $\vec{q}\to\vec{p}$, $\vec{v}\to -\vec{v}$, $\tau\leftrightarrow \tau'$. We obtain the correlator for the variables of definite helicity in momentum space 
\begin{align}\label{eq:corr_mixed_variables_mom}
&\langle h^\lambda (\vec{l},\,\tau)\,h^{\lambda'}(\vec{l}\,',\,\tau')\rangle\equiv e_{ij,\lambda}(-\hat{l})e_{i'j',\lambda'}(-\hat{l}')\int{d^3y\,d^3y'}e^{2\pi i\vec{l}\cdot \vec{y}+2\pi i\vec{l}'\cdot\vec{y}'}\langle h_{ij}(\vec{y},\,\tau)h_{i'j'}(\vec{y}',\,\tau')\rangle\nonumber\\
&= \delta(\vec{l}+\vec{l}\,')e_{ij,\lambda}(-\hat{l})e_{i'j',\lambda'}(\hat{l})\Bigg[\frac{1}{2}\sum_\sigma\int\frac{d^3k}{4\pi\,k^{3}}\,{e^{2\pi iq(\tau-\tau')}}\,\delta(\vec{q}-\vec{l})\,e_{ij,\sigma}(\hat{k})e_{i'j',\sigma}(-\hat{k})P^{\sigma}(k)\nonumber\\
&\qquad\qquad\qquad\qquad\qquad\qquad\qquad\qquad+ (\vec{q}\to\vec{p},\, \vec{v}\to -\vec{v},\, \tau\leftrightarrow \tau')\Bigg]\,.
\end{align}
Our task is then to eliminate $\vec{k}$ from the last equation, expressing it in terms of $\vec{q}$ only.

Firstly, from $d^3k=\gamma\left(1-\hat{q}\cdot\vec{v}\right)\,d^3q$ and $k=\gamma\left(q-\vec{q}\cdot\vec{v}\right)$, we obtain 
\begin{equation}
\frac{d^3k}{k^3} = \left( 1 + 2 \, \hat{q} \cdot\vec{v} \right) \frac{d^3q}{q^3} + \mathcal{O} \left( v^2 \right) \;. 
\end{equation} 
Secondly,  we decompose the product of the two polarization operators in eq.~(\ref{eq:corr_mixed_variables_mom}) in terms of four 1-index quantities $e_{i,\lambda}$ (see eq.~(\ref{u-v-oneindex} )) and we use the identity (\ref{e-1index-identity}), that we can express as a function of $\hat{q}$ using the relation $\hat{k}=\hat{q}-\vec{v}+\hat{q}\left(\hat{q}\cdot\vec{v}\right) + {\rm O } \left( v^2 \right)$, with $\hat{q}=\hat{l}$ as a consequence of the Dirac delta in eq.~(\ref{eq:corr_mixed_variables_mom}). Using these relations and the property $l_ie_{ij,\lambda}(-\hat{l})=0$, we find that, to first order in $\vec{v}$, the part of eq.~(\ref{eq:corr_mixed_variables_mom}) that depends on the polarization operators does not receive any correction at linear order in~$v$:
\begin{align}
e_{ij,\lambda}(-\hat{l})e_{i'j',\lambda'}(\hat{l})\,e_{ij,\sigma}(\hat{k})e_{i'j',\sigma}(-\hat{k})\,\delta(\vec{q}-\vec{l}) & = e_{ij,\lambda}(-\hat{l})e_{i'j',\lambda'}(\hat{l})\,e_{ij,\sigma}(\hat{l})e_{i'j',\sigma}(-\hat{l})+\mathcal{O}(v^2) \nonumber \\
& =\delta_{\lambda\sigma}\delta_{\lambda\sigma'}+\mathcal{O}(v^2)\,.
\end{align}
Finally we expand $P^\lambda(k)=P^\lambda(\gamma\left(q-\vec{q}\cdot\vec{v}\right)) = P^\lambda(q)-(\vec{q}\cdot\vec{v}) \,P^\lambda{}'(q)+\mathcal{O}(v^2)$.

Using these results, and accounting for both terms in the second line of eq.~(\ref{eq:corr_mixed_variables}), we finally obtain the correlator in the boosted frame 
\begin{align}
\langle h^\lambda(\vec{l},\,\tau)\,h^{\lambda'}(\vec{l}\,',\,\tau')\rangle=&\delta_{\lambda\lambda'}\frac{\delta^{(3)}(\vec{l}+\vec{l}\,')}{4\pi\,l^3}\,\Bigg\{P^\lambda(l)\,\cos[2\pi l(\tau-\tau')]\nonumber\\
&+i(\hat{l}\cdot \vec{v})\,\left[2P^\lambda(l)-l\,P^\lambda{}'(l)\right]\,\sin[2\pi l(\tau-\tau')]\Bigg\}+\mathcal{O}(v^2).
\label{eq:h2pt}
\end{align}
It is worth noting that the dipole contribution vanishes in the equal-time case. This is because $\langle h^\lambda(\vec{l},\,\tau)\,h^{\lambda}(\vec{l}\,',\,\tau')\rangle = \langle h^{\lambda}(\vec{l}\,',\,\tau') h^\lambda(\vec{l},\,\tau)\rangle $, which implies that the correlator is invariant under $\vec{l} \leftrightarrow \vec{l}'$
in the equal time case.

\section{Measuring the SGWB net circular polarization  with LISA}
\label{sec: sec_LISA}

We now discuss how the kinematically induced dipolar anisotropy can be used to measure the net circular polarization of SGWBs with the planar  interferometer LISA.
This was first studied in \cite{Seto:2006hf,Seto:2006dz}, where it was  noticed that a measurement of parity odd SGWB anisotropies
can be
used to detect parity violating effects in gravitational interactions.
 Those works focus on the small frequency limit of the detector response functions, and make use of the properties of
 the detector in such regime, as discussed in \cite{Cutler:1997ta,Prince:2002hp}.
 In our work, we first  systematically   discuss, in Section   \ref{char_LISA}, 
   the general properties  of the instrument response functions under parity symmetry, clarifying the relation between these properties and measurements of circular polarization.
 In Section \ref{SNR_LISA}, using the most up-to-date LISA instrument specifications and performing an analysis over the full
 LISA frequency band, 
 we  re-assess   the evaluation of the  signal-to-noise ratio associated with measurements of the SGWB circular polarization.
  
\subsection{LISA response functions} \label{char_LISA}


The space-based laser interferometer LISA~\cite{Audley:2017drz} will be a constellation of three satellites placed at the vertices  (here placed at the positions $\{\vec{x}_1, \vec{x}_2, \vec{x}_3\}$) of an (approximate) equilateral triangle with side length $L = 2.5 \,$ million kilometers. Each satellite is connected to the other two via laser links, resulting in three virtual Michelson interferometers with an opening angle of 60 degrees, labelled by their respective central node. A passing gravitational wave modifies the relative arm lengths in each of these interferometers, inducing a difference in the travel time of the laser light performing a round trip in the two interferometer arms. This difference in travel time corresponds to a phase shift between the two laser beams returning to the central node, which can be detected in the resulting interference pattern. 


The time-delay induced by a gravitational wave in the $i$-th interferometer is obtained by integrating along the photon geodesic taking into account the perturbation of the metric due to the gravitational wave. The result can be expressed as a convolution of the gravitational wave with the response function ${\cal Q}^i$ containing the geometry of the detector~\cite{Romano:2016dpx, Maggiore:1900zz, Bartolo:2018qqn},
\begin{align}
\sigma_i \left( t \right) \equiv \frac{\delta t}{t} = \frac{\delta t}{2 L}  = \sum_\lambda \int d^3 k \, h_{\lambda}(\vec{k}, t - L) \,  e_{ab, \lambda}({\hat k}) \, {\cal Q}_{ab}^i(\vec{x}_i, \vec{k}; \{ \hat U_j\} )\,, 
\label{sigma-i}
\end{align}
with 
\begin{equation}
{\cal Q}_{ab}^i(\vec{x}_i, \vec{k}; \{ {\hat U_j} \}) = \frac{1}{4} e^{-2 \pi i \vec{k} \cdot \vec{x}_i} \left[ {\cal T}(kL, \hat{ k} \cdot \hat {U}_{i})  \, \hat {U}_{i}^a {\hat U}_{i}^b  - {\cal T}(kL,- \hat{k} \cdot \hat {U}_{i+2}) \, \hat{U}_{i+2}^a \hat{U}_{i+2}^b \right] \,, 
\label{calQ}
\end{equation} 
where ${\hat U}_i \equiv \frac{\vec{ x}_{i + 1} - \vec {x}_i}{L}$ is the unit vector in the direction of the arm that goes from the satellite $\vec{x}_i$ to the satellite $\vec{x}_{i+1}$. All indices $\{i, i + 1, \dots\}$ in eq.~(\ref{calQ}) are understood to be modulo 3.

The detector transfer function ${\cal T}$ is given by 
\begin{align}
 {\cal T}(k L, {\hat k} \cdot {\hat U}_i) \equiv e^{- \pi \,i\, k\, L [1 + {\hat k} \cdot {\hat U}_i  ]} \, \text{sinc} \left[  \pi \, k \, L \left(  1 - {\hat k} \cdot {\hat U}_i \right) \right] + e^{ \pi\, i\, k\, L [1 - {\hat k} \cdot {\hat U}_i ]} \,  \text{sinc} \left[  \pi \, k \, L \left(  1 + {\hat k} \cdot {\hat U}_i \right) \right] \,,
\label{calT}
\end{align}
which reduces to ${\cal T} \simeq 2$ for $k L \ll 1$. 

Performing linear combinations of the interferometers $\vec{x}_i$ we can construct the Time Delay Interferometry (TDI) LISA  channels $\{A, E, T\}$ \cite{Adams:2010vc}
\begin{align}
 \Sigma_A \equiv \frac{1}{3}(2 \sigma_X - \sigma_Y - \sigma_Z) \,, \quad 
 \Sigma_E \equiv \frac{1}{\sqrt{3}} ( \sigma_Z - \sigma_Y) \,, \quad 
 \Sigma_T \equiv \frac{1}{3}( \sigma_X + \sigma_Y + \sigma_Z) \,.
 \label{eq:AET}
\end{align}
For an isotropic background, we can exploit the symmetry under the exchange of the vertices of the equilateral triangle to see that all self correlators among $\sigma_X ,\, \sigma_Y ,\, \sigma_Z$ are equal to each other, as are all cross correlations. This in particular implies $\left\langle \Sigma_A  \Sigma_A \right\rangle = \left\langle \Sigma_E  \Sigma_E \right\rangle $, while the cross correlations among $\Sigma_A$, $\Sigma_E$ and $\Sigma_T$ vanish. As we will see explicitly below, these statements do not apply to anisotropic components of the SGWB.

The signal induced by a passing gravitational wave in the channels $O = \{ A, E, T \}$ is 
\begin{align}
\Sigma_O(t)  =   \sum_\lambda \int d^3 k \, h_{\lambda}(\vec{k}, t - L) \,  e_{ab, \lambda}({\hat k}) \, {\cal Q}_{ab}^O(\vec{k}; \{ {\hat x_j} \})\,,
 \label{eq:signal_response}
\end{align}
with ${\cal Q}_{ab}^O(\vec{k}; \{ {\hat x_j} \}) =  \sum_i c^O_i  {\cal Q}_{ab}^i(\vec{x}_i, \vec{k}; \{ {\hat U_j} \})$, where the matrix $c$ is given by
\begin{align}
 c = \begin{pmatrix}
      \frac{2}{3} & - \frac{1}{3} & - \frac{1}{3} \\
      0 & - \frac{1}{\sqrt{3}} & \frac{1}{\sqrt{3}} \\
      \frac{1}{3} & \frac{1}{3} & \frac{1}{3} 
     \end{pmatrix} \,.
\end{align}
For more details on the derivation and notation, see Ref.~\cite{Bartolo:2018qqn}.

\subsubsection{Response function to the SGWB monopole and dipole components}

Combining eq.~\eqref{eq:h2pt} and \eqref{eq:signal_response} yields the two-point correlation function in the time domain,
\begin{align}
\left \langle \Sigma_O(t) \Sigma_{O'}(t') \right \rangle  =  \frac{1}{4} \sum_\lambda \int \frac{dk}{k} &  \left[ {\cal M}_{O O'}^\lambda(k) P_\lambda(k) \cos\left[2\pi k(t - t') \right]  \right. \nonumber \\
 & +\left. v \, {\cal D}^\lambda_{O O'} \left( 2 P_\lambda (k) - k P'_\lambda (k) \right)  \sin\left[2\pi k(t - t') \right]  \right] \;, 
 \label{eq:2pt-time}
\end{align}
where we have introduced the \textit{monopole} and \textit{dipole response functions}
\begin{align}
 {\cal M}_{O O'}^\lambda \left( k \right) & \equiv 4 \, \int \frac{d \Omega_{\hat k}}{4 \pi} \;  e_{ab, \lambda}({\hat k})  e_{a'b', \lambda}(- {\hat k})  {\cal Q}_{ab}^O(\vec{k}) \,{\cal Q}_{a'b'}^{O'}(- \vec{k})  \,,  \label{eq:response_monopole}\\
 {\cal D}_{O O'}^\lambda \left( k,  \hat{v} \cdot  {\hat n} \right) & \equiv  4 i   \int \frac{d \Omega_{\hat k}}{4 \pi} \,  e_{ab, \lambda}( {\hat k})  e_{a'b', \lambda}(- {\hat k})  {\cal Q}_{ab}^O(\vec{k})\, {\cal Q}_{a'b'}^{O'}(- \vec{k}) \, {\hat k} \cdot {\hat v} \,, \label{eq:response_dipole}
\end{align}
where ${\hat n}$ is the normal to the plane of LISA, that, for definiteness, we take it to be oriented upwards for an observer for whom the vertices labeled as $\vec{x}_1 ,\, \vec{x}_2 ,\, \vec{x}_3$ follow one another in the anti-clockwise direction. 

The two response functions satisfy the following properties 
\begin{enumerate}
\item  ${\cal M}_{O O'}^\lambda$ and  ${\cal D}_{O O'}^\lambda$ are real,
\item  ${\cal M}_{O O'}^\lambda$ does not depend on the orientation of the detector;   ${\cal D}_{O O'}^\lambda$ depends on the direction of the detector only through the cosine of the angle between  ${\hat n}$ and $ {\hat v}$,
\item   ${\cal M}_{O O'}^\lambda \rightarrow {\cal M}_{O O'}^\lambda \;\;,\;\;   
{\cal D}_{O O'}^\lambda \rightarrow - {\cal D}_{O O'}^\lambda \;$ if $\;\vec{v} \rightarrow - \vec{v}$, 
\item   ${\cal M}_{O O'}^R \left( k \right) = {\cal M}_{O O'}^L \left( k \right) \;\;,\;\;  {\cal D}_{O O'}^R  \left( k,  {\hat v} \cdot  {\hat n} \right)  = - {\cal D}_{O O'}^L   \left( k,  {\hat v} \cdot  {\hat n} \right) $, 
\end{enumerate} 
which we now prove. 

The first property immediately follows from the fact that ${\cal Q}_{ab}^O(- \vec{k}) = ({\cal Q}_{ab}^O(\vec{k}))^*$, and identically for 
the GW polarization operators. 

The second property is a consequence of statistical isotropy of the monopole, and of the statistical isotropy of the dipole under rotations that preserve the direction of ${\vec{v}}$. Let us verify that the above relations ensure these properties. We start by noting that the transfer function ${\cal T}$ depends on ${\hat k}$ only through ${\hat k} \cdot {\hat U}_i$. The argument in the exponential pre-factor in ${\cal Q}^i$ can be expressed as $2 \pi i k {\hat k}\cdot (\vec{x}_0 +  (\vec{x}_i - \vec{ x}_0))$ with $\vec{x}_0$ denoting the center of the equilateral triangle formed by the three satellites. The factor $\exp(2 \pi i k {\hat k} \cdot \vec{x}_0)$ is thus universal to all ${\cal Q}^i$ and drops out in the dipole response function due to the property ${\cal Q}_{ab}^O(- \vec{k}) = ({\cal Q}_{ab}^O({\vec k}))^*$. The remaining factor can also be written as a scalar product between ${\hat k}$ and the direction of the LISA arms. For instance, for $i=1$, we have 
\begin{equation}
{\hat k}\cdot \left( \vec{x}_1 - \vec{x}_0 \right) = 
{\hat k}\cdot \left( \vec{x}_1 - \frac{\vec{x}_1 + \vec{x}_2 + \vec{x}_3}{3}  \right) = 
\frac{{\hat k}\cdot \left( {\hat U}_3  - {\hat U}_2 \right)}{3} \,. 
\end{equation} 
and analogously for $i=2,3$. Therefore, 
\begin{equation}
{\cal Q}_{ab}^O(\vec{k}) {\cal Q}_{a'b'}^{O'}(- \vec{k}) = {\rm function  \; of } \;\; {\hat k}\cdot {\hat U}_1 \;,\;\;  {\hat k}\cdot {\hat U}_2 \;,\;\; {\rm and \; of } \;\;  {\hat k}\cdot {\hat U}_3 \,. 
\label{propertyQQ}
\end{equation} 
As a consequence, any rotation of the LISA instrument (that for this discussion we consider as a rigid equilateral triangle) can be ``compensated'' by a rotation of ${\hat k}$. The rotation of ${\hat k}$ does not change the monopole response function (\ref{eq:response_monopole}), as this is just the integration variable. It follows that every orientation of the instrument results in the same value for the monopole response function. In the case of the dipole response function, any change of 
the orientation of the instrument can be ``compensated'' by a rotation of  ${\hat k}$ and of $\vec{v}$, 
(since also the last factor must be unchanged). 
Again, since  ${\hat k}$ is simply an internal variable, it follows that the dipole response function does not change if we rotate both the instrument and $\vec{v}$. If we now consider a rotation around the direction of $\vec{v}$, we then see that the dipole response function~\eqref{eq:response_dipole} is unchanged for rotations of the instrument that do not change the angle between $\vec{v}$ and the normal to the plane of the instrument. Therefore, it depends on the orientation of the instrument and of the dipole only through the product   ${\hat v} \cdot  {\hat n}$. More specifically, if we consider a coordinate system in which ${\hat n}$ is directed along the $z-$axis, we see that the last factor in eq.~(\ref{eq:response_dipole}) factorizes a cosine of this angle. 

The third property follows immediately from the properties that we just proved, and from the definition of the response functions. 

The fourth property will be essential for our aim of measuring the SGWB circular polarization. 
To prove it, let us consider a mirror transformation with respect to the plane of the detector. Under this transformation, the component of a vector along ${\hat n}$ (that we denote as $\perp$) changes sign, while the component of the vector on the plane of the detector (that we denote as $//$) remains invariant. Therefore, the product ${\cal Q}_{ab}^O(\vec{k}) {\cal Q}_{a'b'}^{O'}(- \vec{k})$ is invariant under this symmetry, due to (\ref{propertyQQ}). As seen from eq.~(\ref{calQ}), only the components of $e_{ab,\lambda}$ along the plane of the detector contribute to the response functions. One can verify by direct inspection (by using the explicit form of eq.~(\ref{uv-choice})) that these components are unchanged if we perform this mirror transformation and we simultaneously change the GW chirality. Namely, $e_{ab,\lambda} Q_{ab}^i \left( k_{//} ,\, k_\perp \right) = e_{ab,-\lambda} Q_{ab}^i \left( k_{//} ,\, -k_\perp \right)$, as we already proved in~\cite{Bartolo:2018qqn}. Under the mirror transformation, $v_\perp$ changes sign. Therefore, the integrand of the monopole response function is unchanged if we perform this mirror symmetry, and we flip the two helicities, while the integrand of the dipole response function changes sign under the same transformations. The change of 
${\vec k}$ can be then ``undone'' by a change of the integration variable. This implies that the monopole response function is invariant when we flip the two helicities, while the dipole response function changes sign. 
\begin{figure}[t]
\centering
 \includegraphics[width = 0.6 \textwidth]{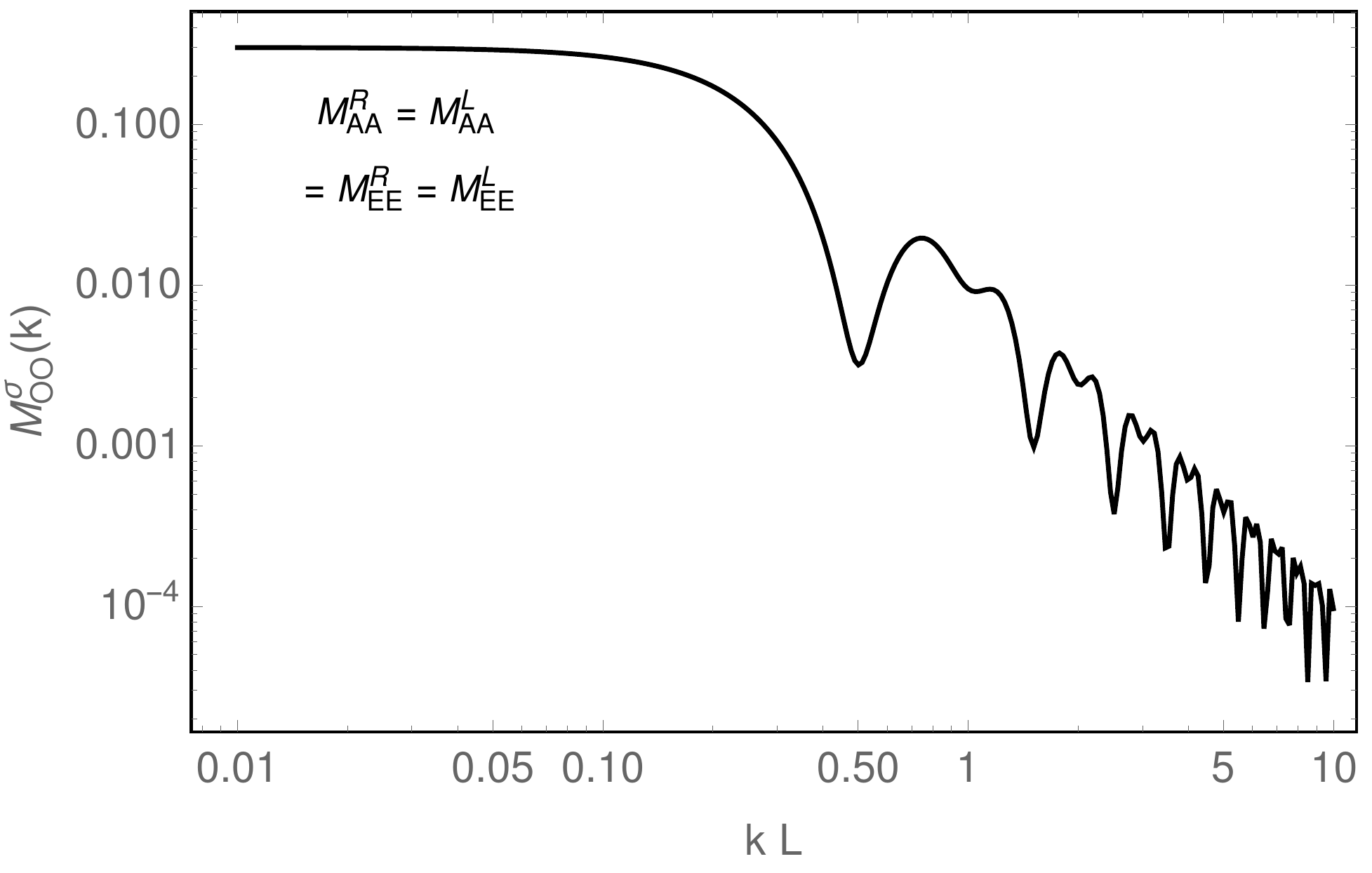}
 \caption{\it Monopole response function. It vanishes in the $AE$ cross-correlation channel while is identical in the $AA$ and $EE$ auto-correlation channels and is insensitive to the chirality of the SGWB.  }
 \label{fig:monopole_response}
\end{figure}

Having proved the above properties, let us now consider a re-labeling of two satellites, say ${\vec x}_2 \leftrightarrow {\vec x}_3$. 
We see from the definitions (\ref{eq:AET}) that the $\Sigma_A$ measurement is invariant under this re-labeling, while $\Sigma_E$ changes sign. 
Therefore, the self-correlators $\left\langle \Sigma_A  \Sigma_A \right\rangle$ and $\left\langle \Sigma_E  \Sigma_E \right\rangle$ are even under the re-labeling, while the cross-correlator  $\left\langle \Sigma_A  \Sigma_E \right\rangle$ is odd. The re-labeling has the effect of inverting the direction of the normal to the plane of the instrument, as we have defined it below eq.~(\ref{eq:response_dipole}). Due to the property (2.) demonstrated above, the monopole response function is invariant under this inversion, while the dipole response function changes sign. Therefore 
\begin{equation}
{\cal M}^\lambda_{AE}  = 0 \,, \qquad 
{\cal D}^\lambda_{AA} = {\cal D}^\lambda_{EE} = 0 \,.
\end{equation} 
These relations can be immediately verified by a direct evaluation of eqs.~\eqref{eq:response_monopole} and \eqref{eq:response_dipole}.\footnote{Similarly, re-labeling of the tensor indices in eq.~\eqref{eq:response_dipole} while simultaneously flipping ${\hat k} \mapsto - {\hat k}$ yields ${\cal D}_{AE}^\lambda = - {\cal D}_{EA}^\lambda$.  Consequently, since $\langle \Sigma_A(t) \Sigma_E(t') \rangle = \langle \Sigma_E(t') \Sigma_A(t) \rangle$, we conclude that the dipole contribution to  $\langle \Sigma_A(t) \Sigma_E(t') \rangle$ must be odd under the exchange $t \leftrightarrow t'$, as reflected by the sine function in eq.~\eqref{eq:2pt-time}. On the contrary, the auto-correlations $\langle \Sigma_A(t) \Sigma_A(t') \rangle$ and $\langle \Sigma_E(t) \Sigma_E(t') \rangle$ trivially have to be even under $t \leftrightarrow t'$.}
\begin{figure}[t]
\centering
 \includegraphics[width = 0.6 \textwidth]{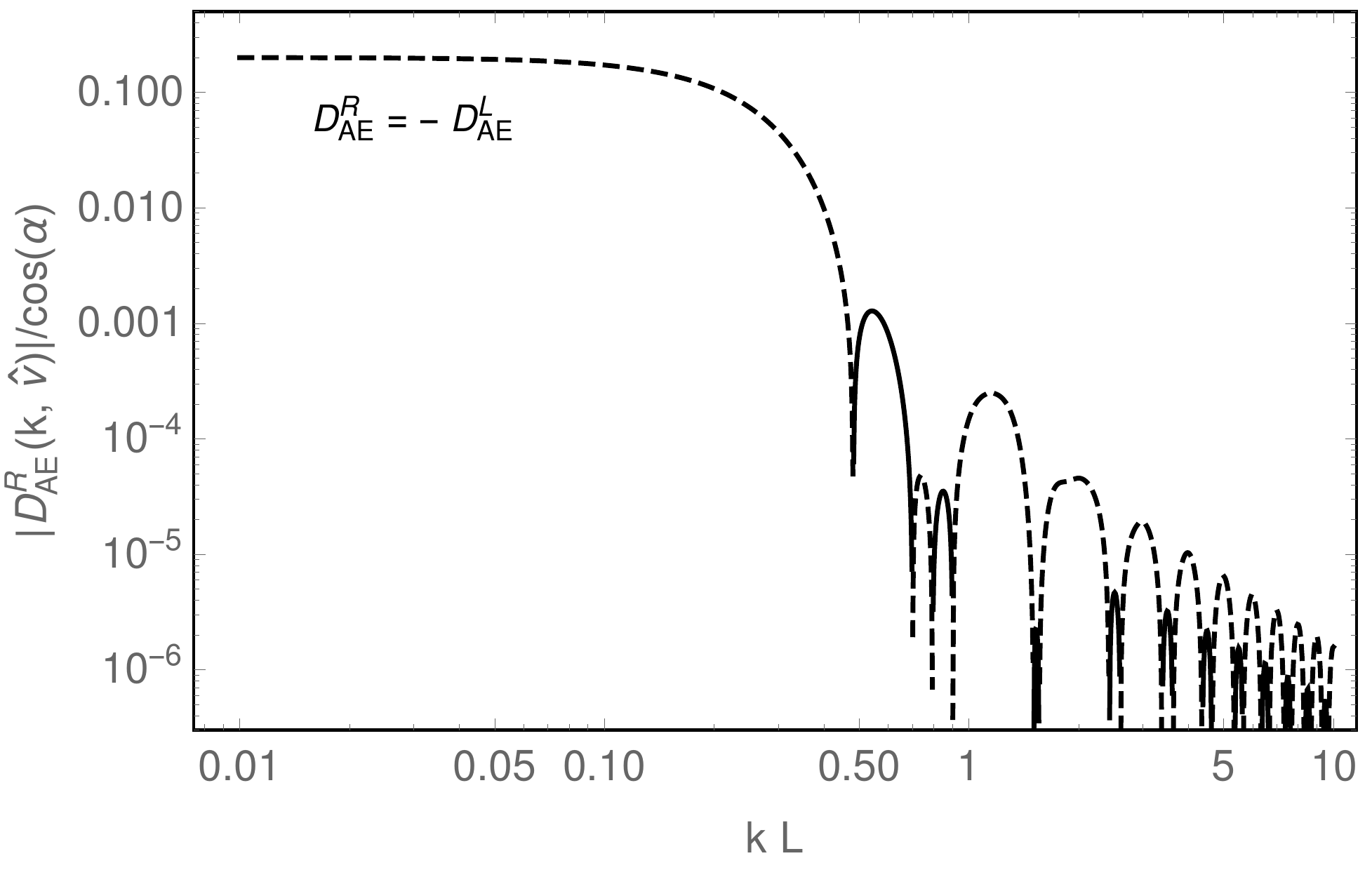}
 \caption{\it
Absolute value of the dipole response function. The dashed (solid) line indicates positive (negative) values for ${\cal D}_{AE}^R = - {\cal D}_{AE}^L$. The angle $\alpha$ denotes the angle between the orientation of the dipole ${\hat v}$ and the plane of the detector. 
 }
 \label{fig:dipole_response}
\end{figure}
In Figs.~\ref{fig:monopole_response} and \ref{fig:dipole_response} we depict the monopole response functions for the $AA$ and $EE$ channel as well as the dipole response function for the $AE$ channel. We recall (property (4.) above) that the dipole response function is odd under a flip of helicity, $\lambda \mapsto - \lambda$, again reflecting that the dipole response function is parity odd. In particular, due the summation over helicity, the total two-point function $\langle \Sigma_A \Sigma_E \rangle$ will only be non-zero if the stochastic background is chiral, i.e.\ if $P_\lambda(k) \neq P_{- \lambda}(k)$.

An important consequence of this is that one should be careful in assuming that a nonvanishing value for $\left\langle \Sigma_A \Sigma_E \right\rangle$ would be due only to noise. As we proved above, this cross-correlator vanishes in presence of the monopole only, and one might be tempted to use any non-zero result as a toll for noise characterization. We have shown that this quantity is actually non-vanishing if the SGWB has a net polarization.

\subsubsection{Dipole antenna pattern}

\begin{figure}
\center
 \includegraphics[width = 0.22 \textwidth]{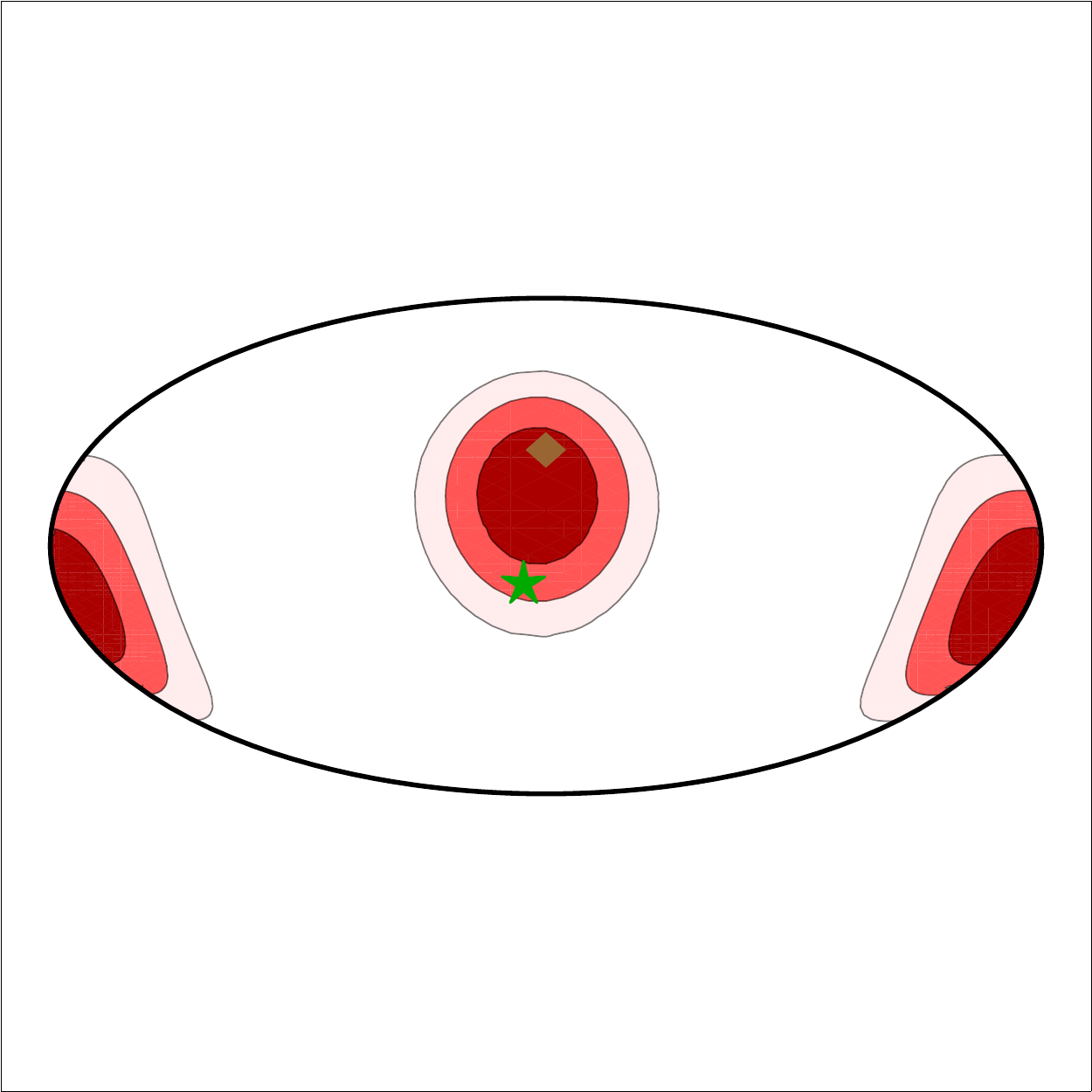} 
  \includegraphics[width = 0.22 \textwidth]{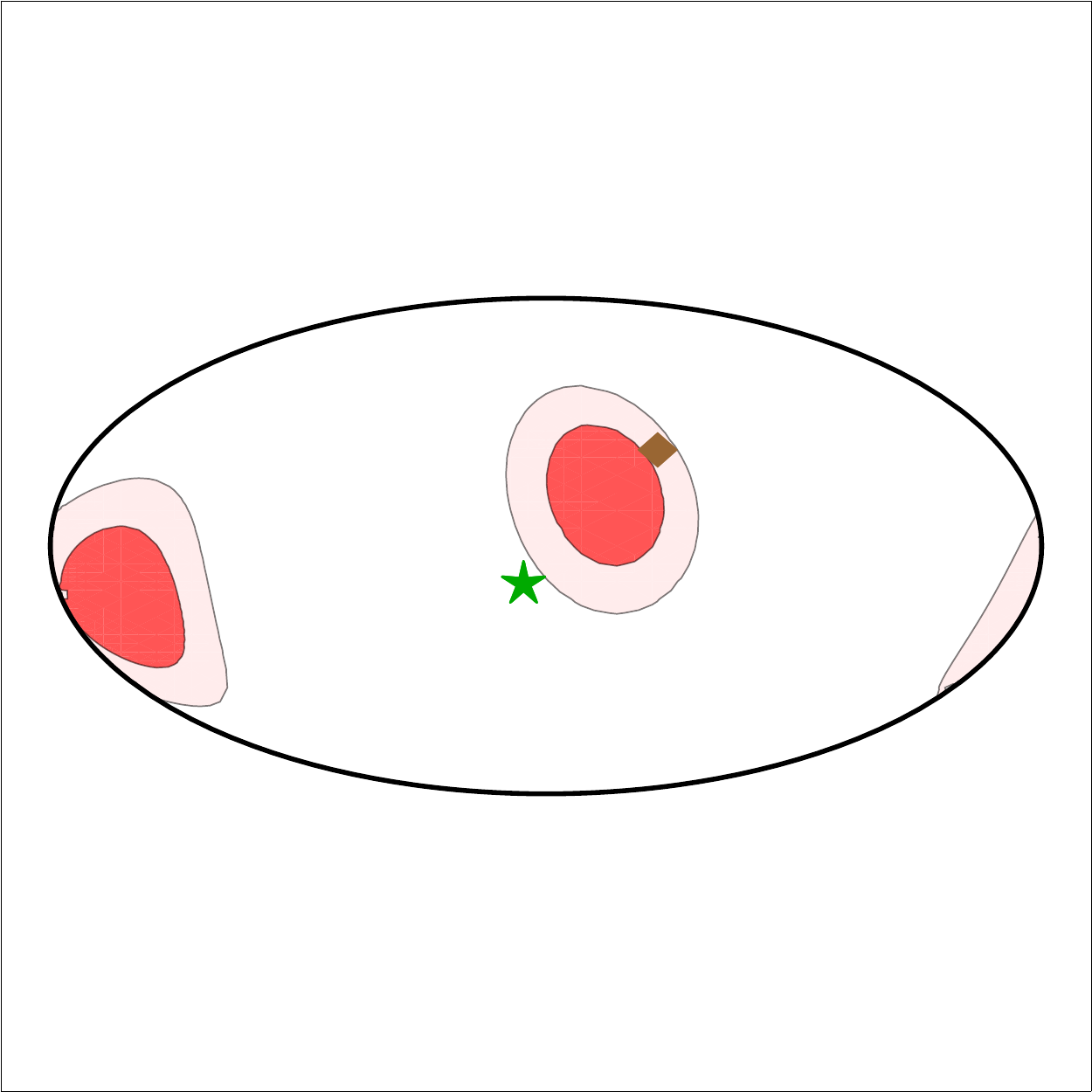} 
    \includegraphics[width = 0.22 \textwidth]{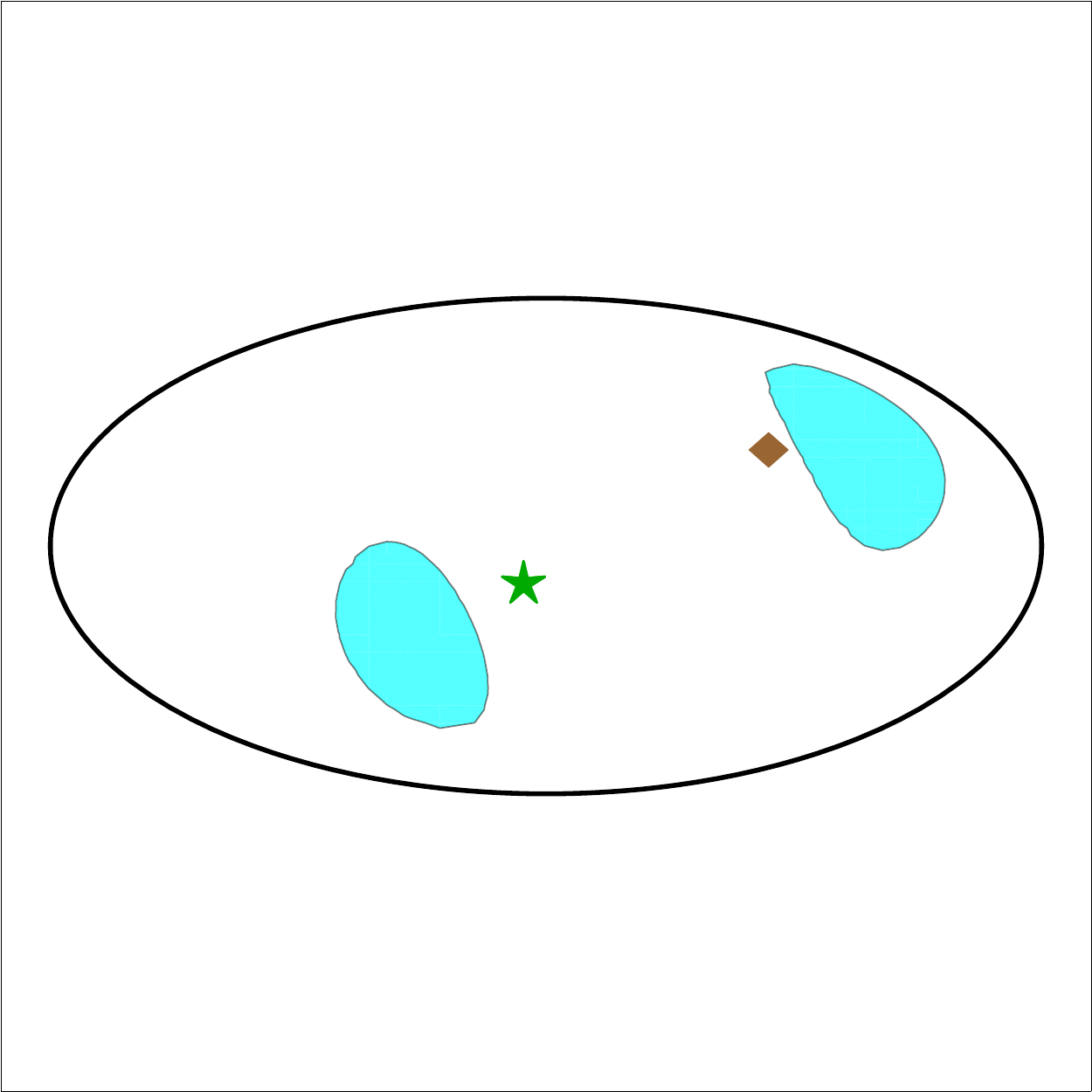} 
      \includegraphics[width = 0.22 \textwidth]{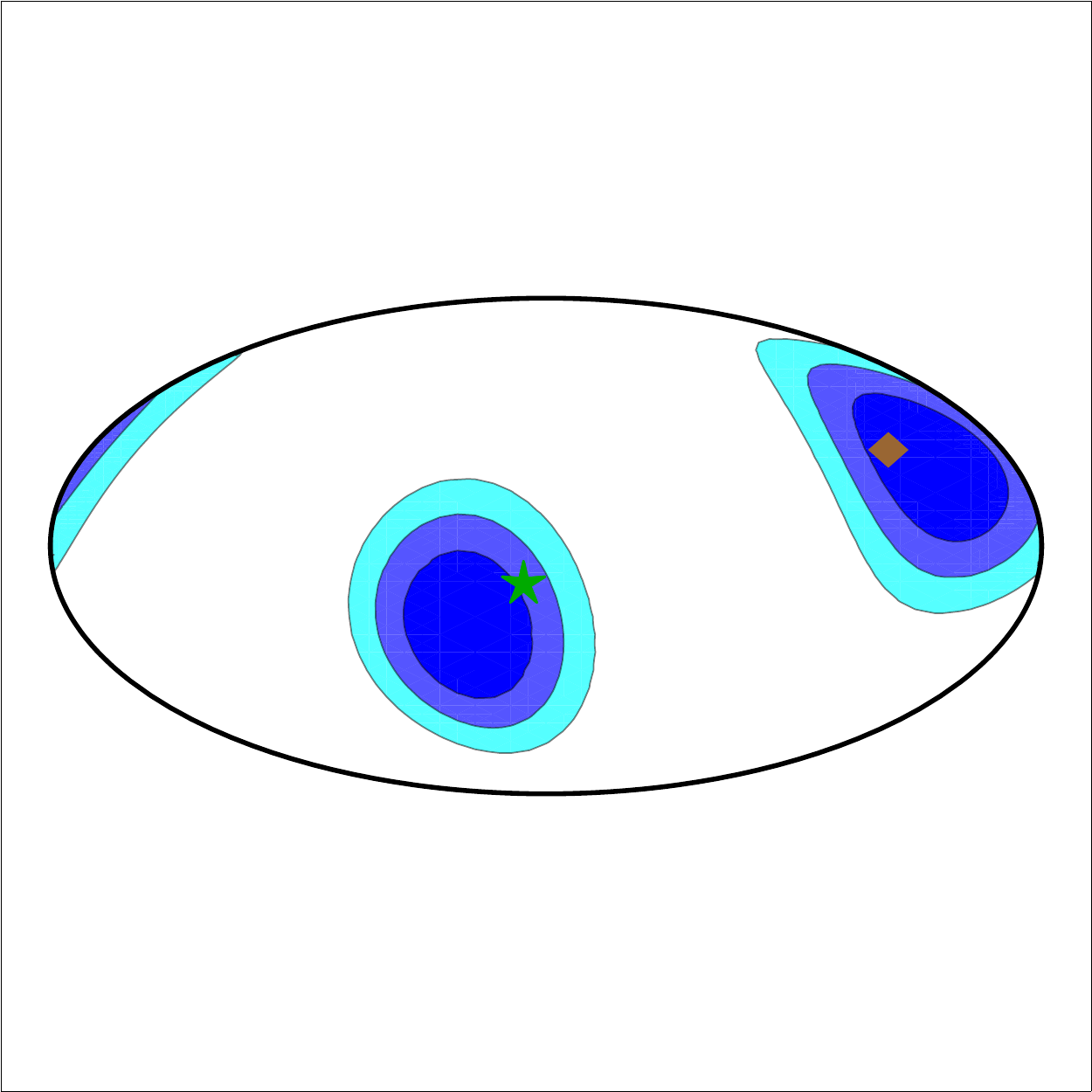}
      
      \vspace{1mm}
      
  \hspace{0.5mm} \includegraphics[width = 0.22 \textwidth]{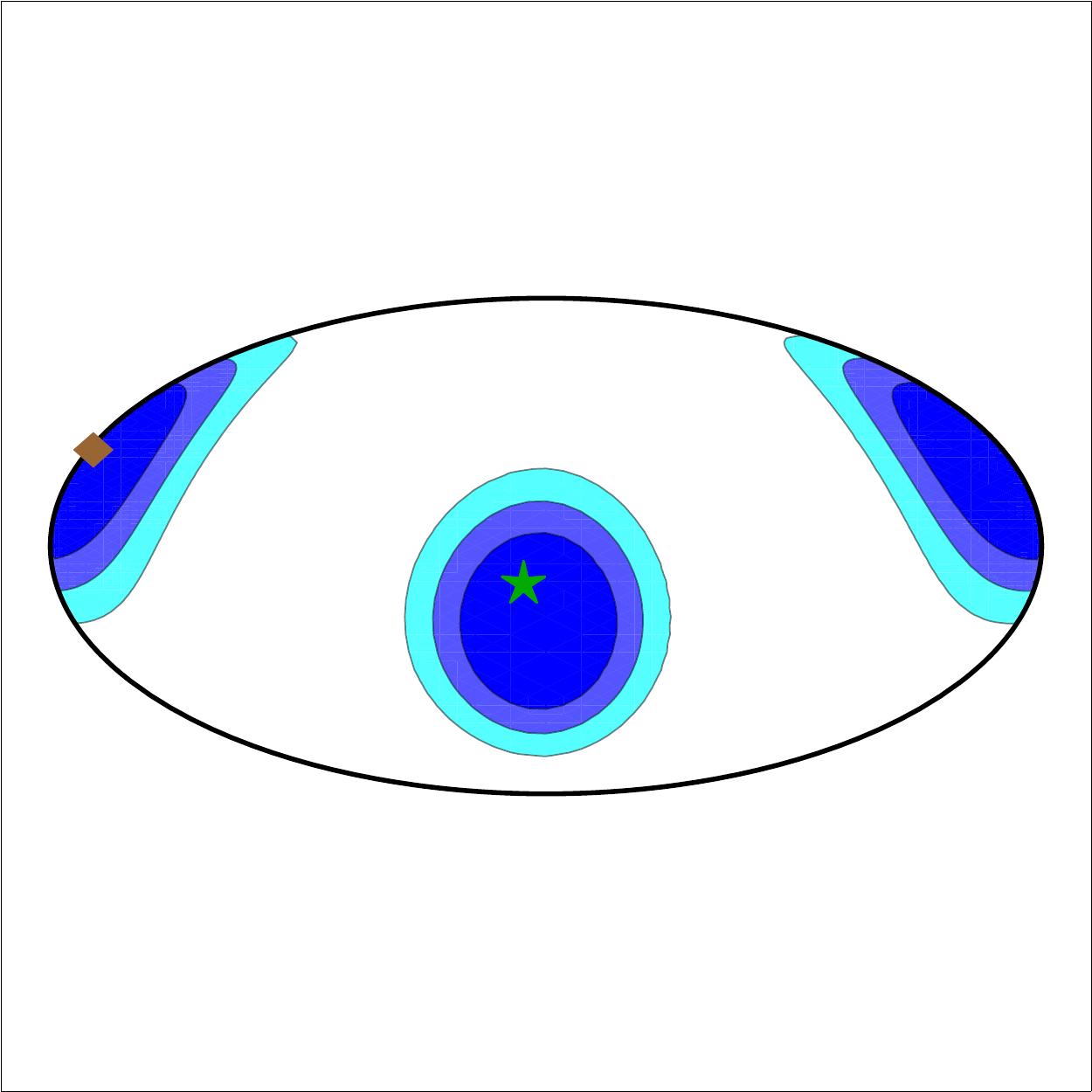} 
   \includegraphics[width = 0.22 \textwidth]{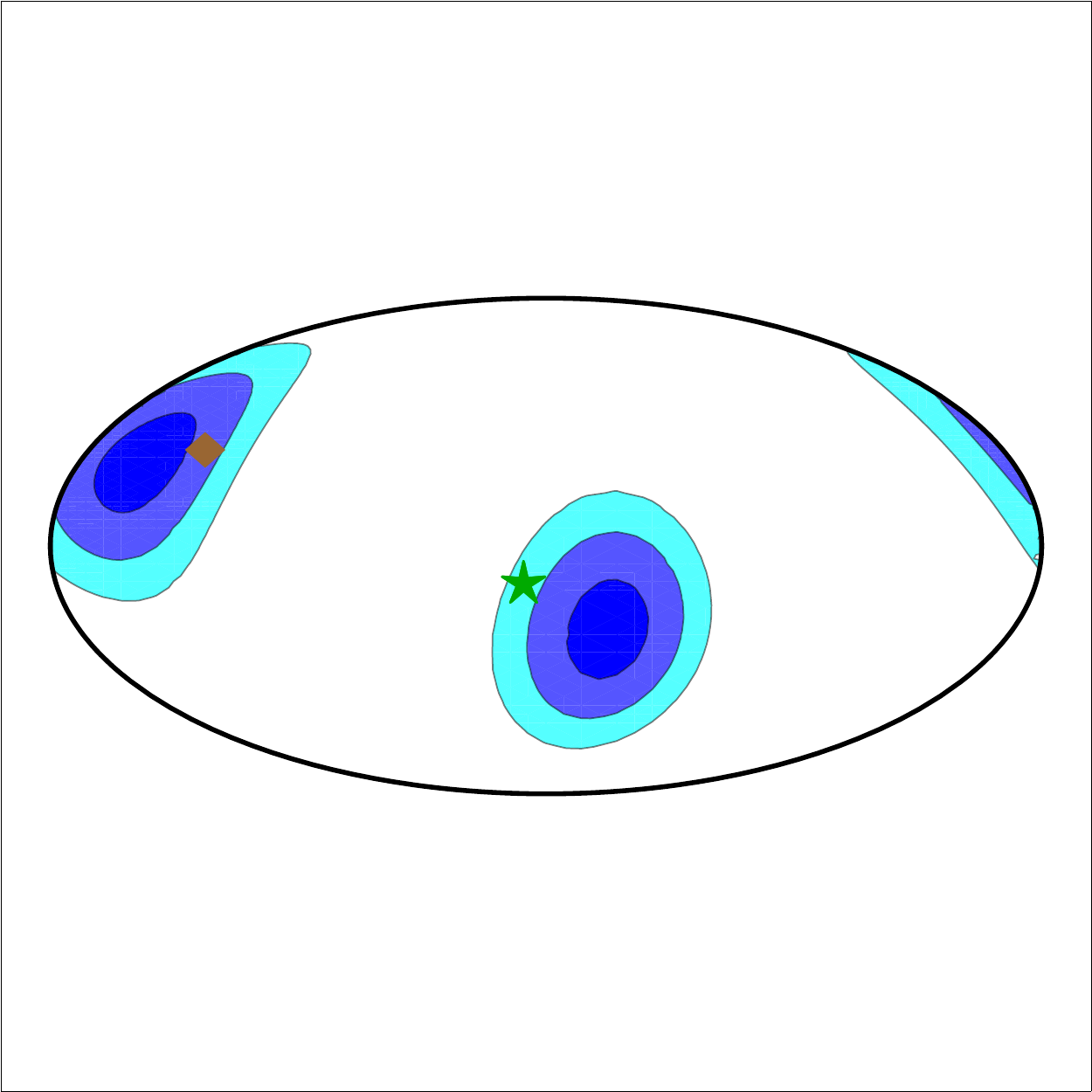} 
    \includegraphics[width = 0.22 \textwidth]{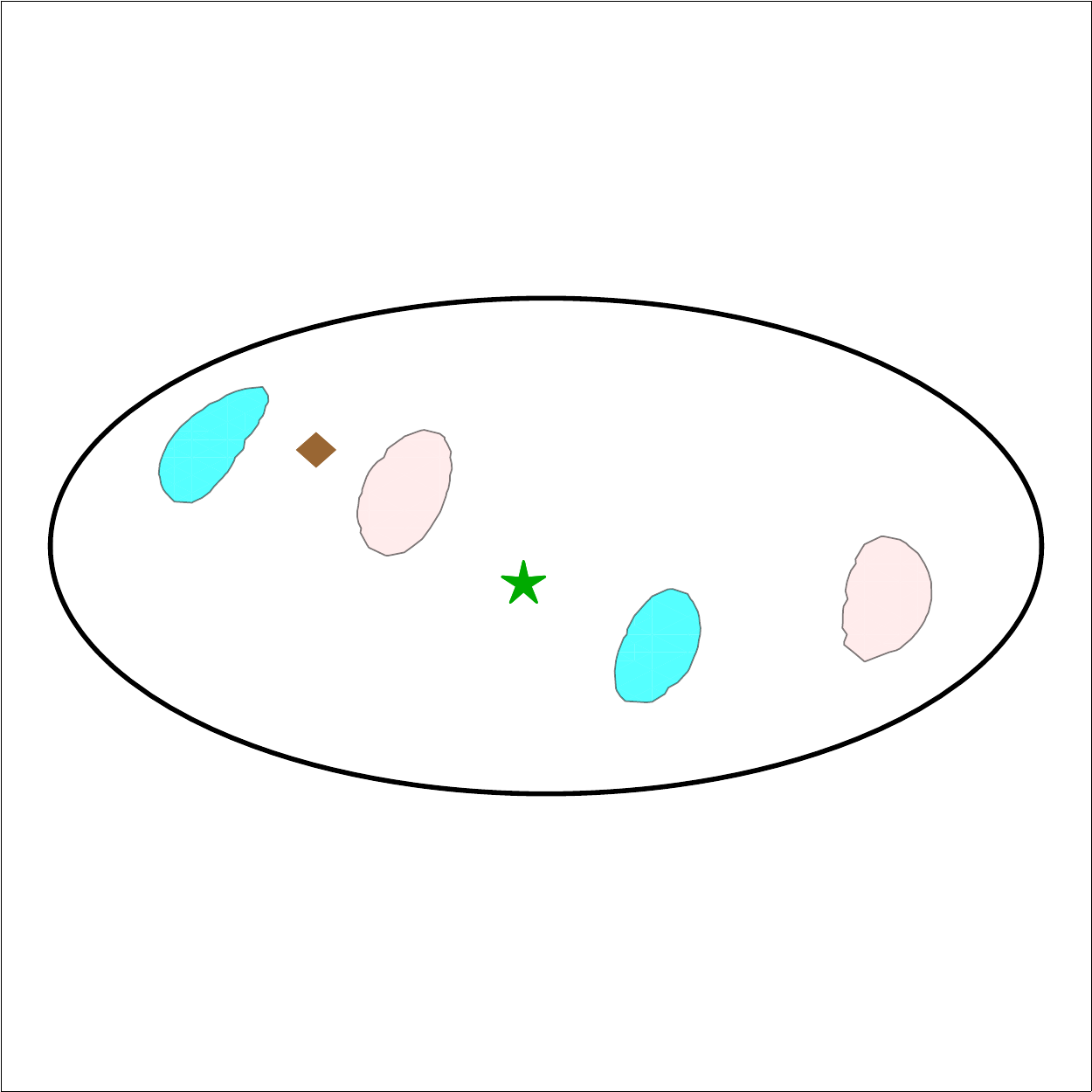}
        \includegraphics[width = 0.22 \textwidth]{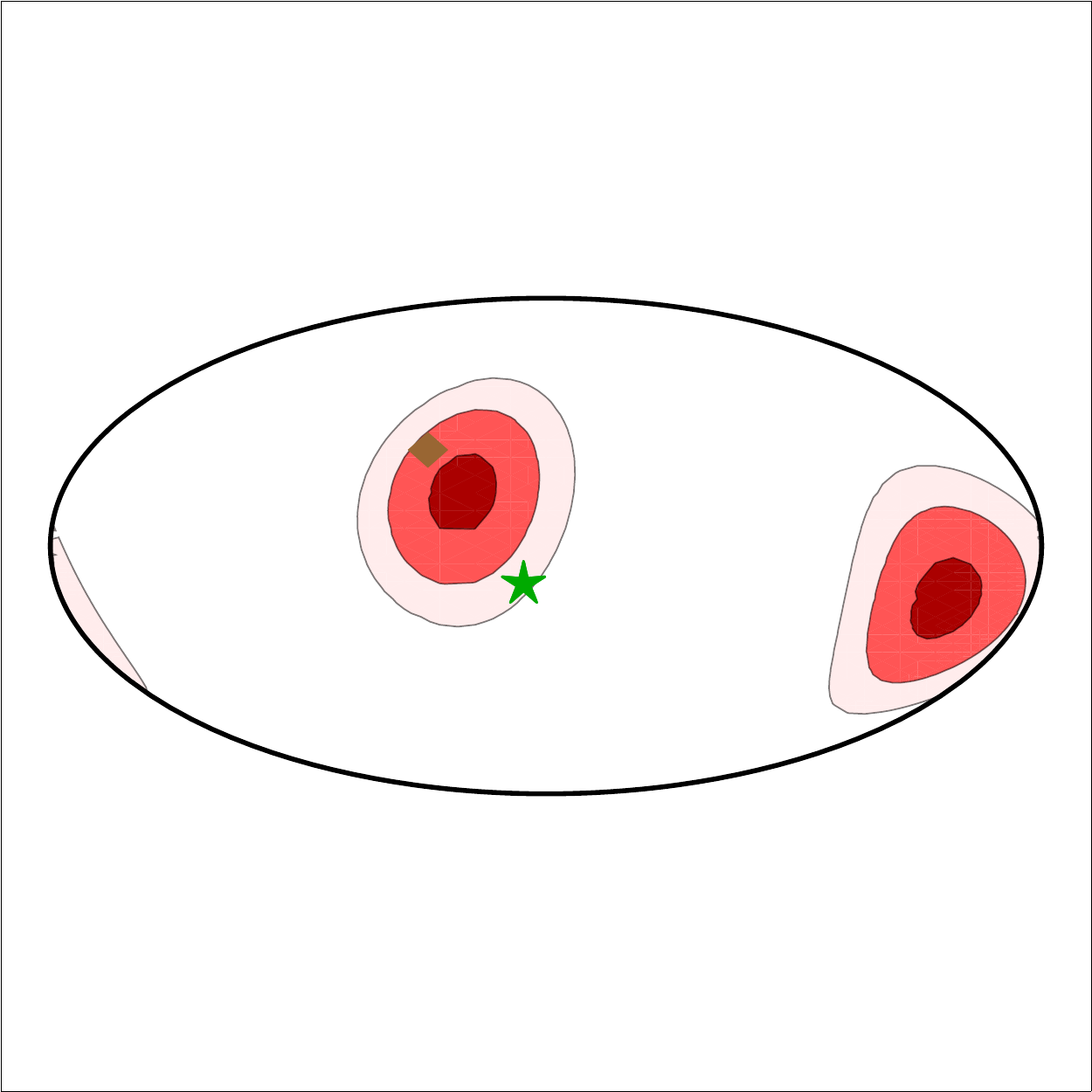}
\caption{\it Evolution of the dipole antenna pattern in ecliptic coordinates induced by the satellite rotation. The plots show the real part of the integrand of ${\cal D}_{AE}^R$ for $f = 10^{-3} \, {\rm Hz}$ and every $1.5$ months. The contour lines are at $0.04, 0.03, 0.02, -0.02, -0.03, -0.04$ (red to blue). The green star denotes the direction of the dipole (assumed to coincide with the CMB dipole), and the brown dot the direction of the LISA normal. } 
\label{fig:antenna-pattern}
\end{figure}

As discussed above, the dipole response function~\eqref{eq:response_dipole} depends only on the angle between the dipole and the normal vector of the detector plane, ${\hat v} \cdot {\hat n}$. The directional sensitivity of the integrand of eq.~\eqref{eq:response_dipole} is more involved, encoding the geometrical sensitivity of the detector to different sky regions, the so-called antenna pattern. The antenna pattern of the monopole response function shows that GW interferometers are most sensitive to GWs arriving orthogonally to the detector plane (see e.g.~\cite{Maggiore:1900zz}). In Fig.~\ref{fig:antenna-pattern} we depict the corresponding dipole antenna pattern, taking into account that, due to the motion of the LISA-plane around the sun, the effective dipole will receive an annual modulation. See Section~\ref{SNR_LISA} for more details about the LISA orbit parametrization.

These antenna patterns give allow for a qualitative understanding  of the resolution of GW detectors to higher order parity odd anisotropies. Moreover, as we will discuss in Sec.~\ref{SNR_LISA}, the expected annual modulation of the dipole response function can be used to optimize the signal-to-noise ratio of this measurement. This is in particular true if the SGWB dipole coincides with the (known) dipole of the CMB.

\subsubsection{Small frequency limit of the response functions} 
\label{subsec:small-kL}

In the small frequency limit, $k \, L \ll 1$, we can Taylor-expand the integrands of eqs.~(\ref{eq:response_monopole}) and (\ref{eq:response_dipole}), and then perform the integrals numerically. We obtain 
\begin{eqnarray}
\label{eq:small_frequency}
&& {\cal M}_{AA}^\lambda \left( k \right) = {\cal M}_{EE}^\lambda \left( k \right) = \frac{3}{10} - \frac{169 \, \pi^2}{420} k^2 \, L^2 
+ \mathcal{O} \left( k^4 L^4 \right) \;\;, \nonumber\\ 
&& {\cal M}_{TT}^\lambda \left( k \right) = \frac{\pi^6}{189} \, k^6 \, L^6 + \mathcal{O} \left( k^8 L^8 \right) \;\;, \nonumber\\ 
&&  {\cal D}_{AE}^\lambda \left( k \right) =  \lambda \,  {\hat v} \cdot  {\hat n} \left[ \frac{1}{5} - \frac{253 \, \pi^2}{840} \, k^2 \, L^2 + \mathcal{O}\left( k^4 L^4 \right) \right]. 
\end{eqnarray}

Obviously, these analytical expressions for the small $k L$ expressions satisfy all the properties of the correlators discussed in the previous subsection. We note that the ${\cal M}_{TT}^\lambda$ correlator vanishes at small frequencies. For this reason this channel is sometimes denoted as the ``null-channel'', and it is expected to provide useful information for noise characterization \cite{Adams:2010vc}. 

For LISA, with
\begin{align}
 2 \pi k L = 0.05 \left( \frac{k}{10^{-3}~\text{Hz}} \right) \left( \frac{L}{2.5 \times 10^6 \text{ km}} \right) \,,
\end{align}
this Taylor expansion is only a good approximation for the lower part of the frequency band. In the following, and in particular in Sec.~\ref{SNR_LISA}, we will work with the full response functions, thereby extending the work of Ref.~\cite{Seto:2006hf}. On the other hand, when we turn to the Einstein Telescope in Sec.~\ref{sec:ET}, the small frequency limit will be fully sufficient. 

\subsubsection{Expressing the results in frequency domain}

Performing a Fourier transform on eq.~\eqref{eq:2pt-time} yields the two-point function in the frequency domain,
\begin{align}
\left \langle \Sigma_O(f) \Sigma_{O'}(f') \right \rangle  = & \frac{1}{4} \sum_\sigma \int \frac{dk}{k} \left[ {\cal M}_{O O'}^\sigma(k) P_\sigma(k)  \int dt \int dt' e^{- 2 \pi i (t f + t' f')} \cos\left[2\pi k(t - t') \right]  \right. \nonumber \\
 &  + \left. v \, {\cal D}^\sigma_{O O'} \left( 2 P_\sigma(k) - k P'_\sigma(k) \right)  \int dt \int dt' e^{- 2 \pi i (t f + t' f')}  \sin\left[2\pi k(t - t') \right]  \right] .
 \label{eq:2pt-f}
\end{align}
Here $f$ and $f'$ can take both positive and negative values and the integration boundaries of the time-integrals are $\bar t - \Delta T/2 \leq t,t' \leq \bar t + \Delta T/2$ where $\bar t$ denotes a reference time and $\Delta T$ the typical length of the data streams in the time domain which for LISA is expected to be ${\cal O}$(10 days). Since $T$ is much longer than the inverse of the
frequency range  LISA is more sensitive to
  (which is  of the order of hours), we will set $\Delta T\to \infty$ from now on. We thus obtain 
\begin{align}
\left \langle \Sigma_O(f) \Sigma_{O'}(f') \right \rangle  =  \frac{1}{4} \sum_\sigma \int_0^\infty \frac{dk}{2k}   \left\{  {\cal M}_{O O'}^\sigma(k) P_\sigma(k) \right.   &  \left[ \delta(-  f + k)\, \delta( f' + k) +\delta( f + k)\, \delta(-  f' + k) \right] \nonumber  \\
  \textcolor{white}{ \int \frac{dk}{2k}} - i   v \, {\cal D}_{O O'}^\sigma(k) \left( 2 P_\sigma(k) - k P'_\sigma(k) \right) &  \left[ \delta(-  f + k)\, \delta( f' + k) - \left.  \delta( f + k) \delta(-  f' + k) \right]
 \right\} \,.
 \label{eq:2pt-f-final}
\end{align}
We note that the dipole contribution is odd under $f \leftrightarrow f'$, indicating that the corresponding contribution to the two-point function must vanish for $O = O'$. This is an immediate consequence of the sine function in eq.~\eqref{eq:h2pt} which indicates that the dipole contribution has support only at unequal times, $t \neq t'$.

\subsection{The optimal signal-to-noise ratio for measuring  circular polarization}
\label{SNR_LISA}

Let $\tilde{s}_O(f)$ be the signal registered by LISA in the $O=\{ A,\,E \}$ channels, in frequency space. The signal will be the sum of a physical signal $\Sigma_O(f) = \delta t(f) / 2L$ and of a noise $\tilde{n}_O(f)$:
\begin{align}
\tilde{s}_O(f)=\Sigma_O(f)+\tilde{n}_O(f)\,.
\end{align}
We define a frequency-dependent estimator 
\begin{align}
\hat{\cal F}(f_1,\,f_2)\equiv  W^{AE}(f_1,\,f_2)\,\tilde{s}_{A}(f_1)\,\tilde{s}_{E}(f_2) \,,
\end{align}
where the filter function $W^{AE}(f_1,\,f_2)$ satisfies the reality condition $W^{AE}(f_1,\,f_2)^*=W^{AE}(-f_1,\,-f_2)$. This implies that the frequency integrated estimator $\hat{\cal F}\equiv  \int df_1\,df_2\,W^{AE}(f_1,\,f_2)\,\tilde{s}_{A}(f_1)\,\tilde{s}_{E}(f_2)$ is real, and has expectation value
\begin{align}
&\langle\hat{\cal F}\rangle=\int df_1\,df_2\,W^{AE}(f_1,\,f_2)\langle\tilde{s}_{A}(f_1)\,\tilde{s}_{E}(f_2)\rangle=i\,\int_{-\infty}^\infty df_1\,W^{AE}(f_1,\,-f_1)\,S_s(f_1)\,,
\end{align}
where in the last step we have defined the $AE$ correlator as 
\begin{align}
\langle\tilde{s}_{A}(f_1)\,\tilde{s}_{E}(f_2)\rangle= \left \langle \Sigma_A(f) \Sigma_{E}(f') \right \rangle=i\,\delta(f_1+f_2)\,S_s(f_1)\,,
\end{align}
with $S_s(f)$ real, and where we have assumed that the noises in the $A$ and in the $E$ channels are uncorrelated. Note that since ${\cal M}_{AE}^\sigma=0$, only the second line of eq.~(\ref{eq:2pt-f-final}) contributes to this expression, that implies that $S_s(-f_1)=-S_s(f_1)$.

We next compute the variance of $\hat{\cal F}$ assuming that the signal is noise dominated, with $\langle n_O(f_1)\,n_{O'}(f_2)\rangle=\delta_{OO'}\,P_{n,O}(f)\,\delta(f_1+f_2)$, so that
\begin{align}
&\langle \hat{\cal F}^2\rangle=\int_{-\infty}^\infty df_1\,df_2\,df_3\,df_4\,W^{AE}(f_1,\,f_2)W^{AE}(f_3,\,f_4)\langle\tilde{n}_{A}(f_1)\,\tilde{n}_{E}(f_2)\tilde{n}_{A}(f_3)\,\tilde{n}_{E}(f_4)\rangle\nonumber\\
&=\int_{-\infty}^\infty df_1\,df_2\,W^{AE}(f_1,\,f_2)W^{AE}(f_1,\,f_2)^*\,P_{n,A}(f_1)\,P_{n,E}(f_2)\,.
\end{align}
The signal-to-noise ratio (SNR) is then given by $\langle\hat{\cal F}\rangle/\sqrt{\langle \hat{\cal F}^2\rangle}$. 

To determine the filter function $W^{AE}(f_1,\,f_2)$ we define a noise-weighted scalar product in frequency space  as 
\begin{align}\label{eq:def_scalarprodab}
\left(A,\,B\right)=&\int_{-\infty}^\infty df_1\,df_2 A(f_1,\,f_2)\,B(f_1,\,f_2)^* P_{n,A}(f_1)\,P_{n,E}(f_2)\,,
\end{align}
so that the SNR 
\begin{align}\label{eq:snr_theo}
{\rm SNR}=\frac{\left(W^{AE},\,-i\delta(f_1+f_2)\,\frac{S_{s}(f_1)}{P_{n,A}(f_1)\,P_{n,E}(f_2)}\right)}{\sqrt{\left(W^{AE},\,W^{AE}\right)}}\,,
\end{align}
is maximized for $W^{AE}(f_1,\,f_2)\propto -i\,\delta(f_1+f_2)\,\frac{S_{s}(f_1)}{P_A^n(f_1)\,P_E^n(f_2)}$. For this optimal estimator, the SNR is thus given by
\begin{align}\label{eq:snr_general}
{\rm {SNR}}=\left[T\int_{-\infty}^\infty df_1\frac{S_{s}(f_1)^2}{P_A^n(f_1)\,P_E^n(-f_1)}\right]^{1/2}=\left[2\,T\int_{0}^\infty df\frac{S_{s}(|f|)^2}{P_A^n(|f|)\,P_E^n(|f|)}\right]^{1/2}\,,
\end{align}
where $T$ is the total duration of the measurement. 

Next, we write explicitly $S_s(f)$ using the response function ${\cal D}^\lambda_{AE}(k)=\lambda\,D(kL)\,\cos\alpha$, where the function $D(x)$ is plotted in Figure~\ref{fig:dipole_response}. The quantity ${\cal D}^\lambda_{AE}(k)$ (and, consequently, the expectation value for the signal $S_s(f)$) depends on time through the angle $\alpha$ between the direction of the motion of  the solar system and the normal to LISA's plane that rotates as the detector orbits the Sun. As a consequence we will write the signal from now on as $S_s(f,\,T)\propto \cos\alpha(T)$ and, when computing the SNR, we will replace the factor $T$ in eq.~(\ref{eq:snr_general}) with an integral over $dT$, assuming that the typical timescale on which Fourier transforms are computed is much shorter than the month-long timescale on which $\alpha(T)$ changes significantly.

We thus obtain
\begin{align}
S_s(f,\,T)=\frac{3\,v\,H_0^2}{2\,\pi^2\,f^3}\,D(|f|L)\left(\sum_\lambda \lambda\,\Omega_{GW}^\lambda\right)\,\cos\alpha(T)\,,
\label{eq:SNR0}
\end{align}
where we have used the relation $P^\lambda(f)=\frac{3H_0^2}{\pi^2 f^2}\Omega^\lambda_{GW}$, and where we assume, to have a  measure of the reach of this observable, that $\Omega_{GW}^\lambda$ is does not depend on frequency within the LISA bandwidth, which implies that the quantity $2 P^\lambda(|f|) - |f|\, P^\lambda{}'(|f|)$ that appears in eq.~(\ref{eq:2pt-f-final}) equals $\frac{12\,H_0^2}{\pi^2 f^2}\Omega^\lambda_{GW}$. It is worth reminding here that  $P^\lambda(f)$ is the gravitational wave power spectrum evaluated at the time of detection, which is different from the primordial gravitational wave power spectrum, and is related to it by the transfer function (see footnote~\ref{ftn:cosmic}).

In order to determine the noise spectral functions $P_{n,A}(f)=P_{n,E}(f)\equiv \frac{2}{3}P_n(f)$ we use the formulae given in~\cite{Cornish:2018dyw,Caprini:2019pxz}, that give
\begin{align}
f\,P_{n}(f)\simeq &\,7\times 10^{-43}\left(\frac{f}{f_*}\right)\left(1+10^{-4}\left(\frac{f_*}{f}\right)^4\right)\nonumber\\
&+2.3\times 10^{-46}\,(1+\cos^2(f/f_*))\,\left(\frac{f_*}{f}\right)^3\left(1+4\times 10^{-4}\left(\frac{f_*}{f}\right)^2\right)\left(1+39\left(\frac{f}{f_*}\right)^4\right)\,,
\end{align}
where $f_*=(2\pi L)^{-1}\simeq .02\,$Hz.

The final expression for the signal-to-noise ratio, for a scale invariant $\Omega_{GW}^\lambda$, is thus
\begin{align}
\label{snr_lisa}
{\rm {SNR}}&=\frac{9\,H_0^2}{4\,\pi^2}\,v\,\left|\sum_\lambda\lambda\,\Omega_{GW}^\lambda\right|\,\left[2\int dT\,\cos^2\alpha(T)\,\int_0^{\infty} \frac{df}{f^4}\frac{D(fL)^2}{(f\,P_n(f))^2}\right]^{1/2}\nonumber\\
&\simeq 8.5\times 10^{13}\,v\,\left|\sum_\lambda\lambda\,\Omega_{GW}^\lambda h^2\right|\,\left[\int_0^{\frac{T}{1\,{\rm {year}}}}\cos^2\alpha(x)\,dx\right]^{1/2} \,.
\end{align}
Next, we have to estimate the integral $\int_0^{\frac{T}{1\,{\rm {year}}}}\cos^2\alpha(x)\,dx$. LISA will be orbiting the Sun with its normal vector at $30^o$ with respect to the ecliptic plane, pointing south~\cite{Audley:2017drz}. Placing the ecliptic on the $xy$ plane, and approximating that the orbit of the Earth with a circle, the unit vector normal to LISA's plane has components
\begin{align}
{\bf n}=\left(\frac{\sqrt{3}}{2}\cos\left(2\pi \frac{t}{1\,{\rm {year}}}\right),\,\frac{\sqrt{3}}{2}\sin\left(2\pi \frac{t}{1\,{\rm {year}}}\right),-\frac{1}{2}\right)\,.
\end{align}
Parametrizing the velocity vector as ${\bf v}=v(\cos\theta_\bv\,\sin\phi_\bv,\,\cos\theta_\bv\,\cos\phi_\bv,\,\sin\theta_\bv)$, we have 
\begin{align}
\cos\alpha={\bf n}\cdot \bv=\frac{\sqrt{3}}{2}\,\sin\left(2\pi \frac{t}{1\,{\rm {year}}}+\phi_\bv\right)\,\cos\theta_\bv-\frac{\sin\theta_\bv}{2}\,.
\end{align}

The integral of $\cos^2\alpha$ over $1$~year gives the result
\begin{align}\label{int1y}
\left[\int_0^{1}\cos^2\alpha(x)\,dx\right]^{1/2}=\frac{\sqrt{5+\cos(2\theta_\bv)}}{4}\,,
\end{align}
that, depending on the value of $\cos\theta_\bv$, ranges between $.5$ and $.61$. The value of the integral over the total time $T$ of observation, which appears in eq \eqref{snr_lisa}, can then be found multiplying 
the result of eq  \eqref{int1y} by $\sqrt{T/({\rm 1\,\, year})}$. 

Thus, approximating $\left[\int_0^{1}\cos^2\alpha(x)\,dx\right]^{1/2}\simeq .5$, the total SNR turns out to be given approximately by 
\begin{align}\label{snr_final}
{\rm {SNR}}&\simeq \left(\frac{v}{10^{-3}}\right)\,\left|\frac{\sum_\lambda\lambda\,\Omega_{GW}^\lambda h^2}{1.4 \cdot 10^{-11}}\right|\,\sqrt{\frac{T}{3\,{\rm {years}}}}\,.
\end{align}
 This is one order of magnitude larger than the estimate obtained in \cite{Seto:2006hf}.

 For definiteness, given that we use a different notation, we  present  in Appendix \ref{app:comparison} 
a detailed comparison among our computation and Seto's results of \cite{Seto:2006hf}.  On the other hand, we stress
that for our analysis we use the most up-to-date LISA instrument specifications, and more complete formulas
valid for the entire frequency band of the interferometer.

\section{Measuring the SGWB net circular polarization with ground-based interferometers}
\label{sec:ground_based}

We now apply the formulas and techniques of the previous section to the case of ground-based interferometers. 
We develop fully analytical, `covariant' formulas for overlap functions, describing correlations among  ground based interferometers in the small antenna limit (condition ({\ref{condition-T=2}) below) \footnote{As customary in the literature, we call overlap functions the response functions for GW experiments that correlate distinct detectors.}.
Our formulas include  the possibility that the SGWB is circularly polarized, do not  rely on special choices of frame (this is why we call them covariant), and apply to any detector shape (not limited to interferometers with orthogonal arms).  When correlating
 distinct ground based interferometers, it is well known that  the SGWB monopole is already sensitive
 to circular polarization (see e.g. \cite{Seto:2007tn,Seto:2008sr,Crowder:2012ik}). We demonstrate this fact 
 in terms of our analytic formulas, discuss the most convenient detector locations for maximizing sensitivity
 to circular polarization, and also include  the kinematically induced dipole in our analysis. 
In the final part of this section we turn to the future ground-based Einstein Telescope. A single instrument of this type will be planar,  and hence measuring the chirality of the SGWB requires taking into accoung the kinematic dipole, as in the analysis for LISA.

\smallskip

Our starting point is given by relations (\ref{eq:response_monopole}) and  (\ref{eq:response_dipole}), which apply  also to pairs of ground-based interferometers (we actually choose a different overall normalization, as we discuss below). In these cases, the fact that the peak sensitivity of these detectors is at a frequency which is small compared to their inverse arm length, results in a crucial simplification, allowing us to obtain fully analytical expressions for the overlap functions. Covariant,  analytical formulas for the unpolarized overlap function to the SGWB monopole ${\cal M}^R \left( k \right) + {\cal M}^L \left( k \right)$ can already be found in the literature \cite{Allen:1997ad,Flanagan:1993ix}. Here for the first time we provide covariant,  analytic expressions for the $\lambda-$dependent terms (contrary to LISA, these terms do not generally vanish, since pairs of detectors located in different locations on the Earth are generally not coplanar). Moreover, for the first time we provide a covariant,  analytic expressions for the  overlap function to the SGWB dipole.  

For ground-based detectors, the crucial simplification arises from the fact that their sensitivity region  satisfies the ``short arm condition'' (referred to as ``small $kL$ limit'' in Sec.~\ref{subsec:small-kL})
\begin{equation}
2 \pi k \, L \simeq 0.0084 \, \frac{k}{100 \, {\rm Hz}} \,  \frac{L}{4 \, {\rm km}} \ll 1 \;\;,
\label{condition-T=2}
\end{equation}  
where we have normalized the frequency $k$ to the region of best sensitivity for the existing and forthcoming detectors, and where we recall that  the arms of the two LIGO sites are $L=4$ km long, while those of Virgo and KAGRA are $L=3$ km long. In this limit, the quantity ${\cal T}$ indroduced in eq.~(\ref{calT}) evaluates to ${\cal T} \to 2$. Using this value, eq.~(\ref{sigma-i}) assumes the simpler form 
\begin{align}
\sigma_i \left( t \right) \equiv \frac{\delta t}{t} =   D_i^{ab} \sum_\lambda \int d^3 k \, h_\lambda \left( {\vec k}, t - L \right) \,  e_{ab, \lambda}({\hat k}) \,  e^{-2 \pi i \vec{k} \cdot \vec{x}_i} \;\;\;,\;\;\; D_i^{ab} \equiv  \frac{{\hat U}_{i}^a {\hat U}_{i}^b  -  {\hat V}_{i}^a {\hat V}_{i}^b}{2} \,, 
\end{align}
where now ${\hat U}_{i}$ and ${\hat V}_{i}$ are the orientations of the arms of the $i-$th detector, that start from the common point located at ${\vec x}_i$. In the following, we refer to this point as to the ``position of the detector'' for brevity. The vectors ${\vec x}_i$,\, ${\hat U}_{i}$,\, and ${\hat V}_{i}$ for the two LIGO detectors, for Virgo, and for KAGRA are given in Appendix \ref{app:detectors}.

Using eq.~(\ref{eq:h2pt}) for the GW correlator, we then obtain an expression identical to (\ref{eq:2pt-time}), namely 
\begin{align}
\left \langle \Sigma_i(t) \Sigma_j(t') \right \rangle  =   \sum_\lambda \int \frac{dk}{k} &  \left[ {\cal M}_{ij}^\lambda(k) P_\lambda(k) \cos\left[2\pi k(t - t') \right]  \right. \nonumber \\
 & - \left. v \, {\cal D}^\lambda_{ij} \left( 2 P_\lambda (k) - k P'_\lambda (k) \right)  \sin\left[2\pi k(t - t') \right]  \right] \;, 
 \label{eq:2pt-time-ground}
\end{align}
with~\footnote{We use a different normalization for the overlap function for ground-based interferometers with respect to the one used for LISA in Sec.~\ref{sec: sec_LISA}, to respect the literature. }
\begin{eqnarray} 
{\cal M}_{ij}^\lambda \left( k \right) &=&  D_i^{ab} \, D_j^{cd} \, \int \frac{d \Omega_k}{4 \pi} \, 
{\rm e}^{-2 \pi i \vec{k} \cdot \left(  \vec{x}_i -  \vec{x}_j \right)} \, 
e_{ab, \lambda}( {\hat k})  e_{cd, \lambda}(- {\hat k}) \nonumber\\ 
{\cal D}_{ij}^\lambda \left( k ,\, {\hat v} \right) &=&   i \, D_i^{ab} \, D_j^{cd} \, \int \frac{d \Omega_k}{4 \pi} \, 
{\rm e}^{-2 \pi i \vec{k}  \cdot \left(  \vec{x}_i -  \vec{x}_j \right)} \, 
e_{ab, \lambda}( {\hat k})  e_{cd, \lambda}(- {\hat k}) \,  {\hat k} \cdot {\hat v} \,. 
\label{eq:response_ground}
\end{eqnarray} 
In Appendix \ref{app:ground-analytic} we compute these expression analytically. Parameterizing the positions of the different detectors as
\begin{equation}
\kappa  \equiv 2 \pi k \vert  \vec{x}_i - \vec{x}_j \vert \;\;,\;\; 
{\hat s}_{ij} \equiv  \frac{\vec{x}_j - \vec{x}_i}{\vert \vec{x}_i - \vec{x}_j \vert} \;, 
\end{equation} 
and introducing the functions
\begin{eqnarray} 
&& f_A \left( \kappa \right) \equiv  \frac{j_1 \left( \kappa \right)}{2 \kappa} + \frac{1-\kappa^2}{2 \kappa^2} j_2 \left( \kappa \right) 
\;\;\;,\;\;\; 
f_B \left( \kappa \right) \equiv \frac{j_1 \left( \kappa \right)}{\kappa}-\frac{5- \kappa^2}{ \kappa^2} \, j_2 \left( \kappa \right) \;\;,  \nonumber\\ 
&& f_C \left( \kappa \right) \equiv  \frac{-7 j_1 \left( \kappa \right)}{4 \kappa}+\frac{35- \kappa^2}{4 \kappa^2} \, j_2 \left( \kappa \right)  \;\;, \nonumber\\ 
&& f_D \left( \kappa \right) \equiv    \frac{j_1 \left( \kappa \right)}{2} -  \frac{j_2 \left( \kappa \right)}{2 \kappa} \;\;\;,\;\;\; 
f_E \left( \kappa \right) \equiv  - \frac{j_1 \left( \kappa \right)}{2} + 5 \frac{j_2 \left( \kappa \right)}{2 \kappa} \;\;, 
\end{eqnarray} 
(where $j_\ell$ are spherical Bessel functions) the overlap function for  the SGWB monopole is 
\begin{eqnarray}
{\cal M}_{ij}^\lambda \left( k \right)  &=&  f_A \left( \kappa \right) \; {\rm tr } \left[ D_i D_j \right] 
+  f_B \left( \kappa \right) \; \left( D_i {\hat s}_{ij} \right)^a \left( D_j {\hat s}_{ij} \right)^a 
+ f_C  \left( \kappa \right) \; \left( D_i {\hat s}_{ij} {\hat s}_{ij} \right)  \left( D_j {\hat s}_{ij} {\hat s}_{ij} \right) \nonumber\\
&+& \lambda \, f_D  \left( \kappa \right) \; \left[ D_i D_j \right]^{ab} \epsilon_{abc} {\hat s}_{ij}^c 
+  \lambda \, f_E  \left( \kappa \right)  \left( D_i {\hat s}_{ij} \right)^a  \left( D_j {\hat s}_{ij} \right)^b  \epsilon_{abc} {\hat s}_{ij}^c \;\;, 
\label{result-M-ground}
\end{eqnarray} 
while that to the SGWB dipole is 
\begin{align} 
{\cal D}_{ij}^\lambda \left( k ,\, {\hat v} \right) &=
f_A' \left( \kappa \right) \, {\hat v}_e {\hat s}_e \left( D_i \, D_j \right)^{aa} \nonumber\\ 
& +  \left[ f_B' \left( \kappa \right) - 2 \, \frac{f_B \left( \kappa \right)}{\kappa} \right] {\hat v}_e {\hat s}_e  
\left( D_i {\hat s}_b \right)^a  \left( D_j {\hat s} \right)^a 
+  \frac{f_B \left( \kappa \right)}{\kappa}  \left[  
\left( D_i {\hat v} \right)^a \, \left( D_j  {\hat s} \right)^a + \left(  D_i {\hat s} \right)^a \, \left( D_j  {\hat v} \right)^a  
\right] \nonumber\\ 
& +  \left[ f_C' \left( \kappa \right) - 4 \, \frac{f_C \left( \kappa \right)}{\kappa} \right] {\hat v}_e {\hat s}_e  \; 
\left( D_i {\hat s} {\hat s} \right) \;  \left( D_j  {\hat s} {\hat s} \right) 
+ 2 \frac{f_C \left( \kappa \right)}{\kappa} \left[ 
\left( D_i {\hat s} {\hat v} \right) \; \left( D_j  {\hat s} {\hat s} \right) + 
\left( D_i {\hat s} {\hat s} \right) \;  \left( D_j  {\hat s} {\hat v} \right) 
\right] \nonumber\\ 
& +  \lambda \left[ f_D' \left( \kappa \right) -  \frac{f_D \left( \kappa \right)}{\kappa} \right] {\hat v}_e {\hat s}_e  
\left( D_i D_j \right)^{ab}  \epsilon_{abc} {\hat s}_c  +  \lambda \frac{f_D \left( \kappa \right)}{\kappa} 
\left( D_i   D_j \right)^{ab}   \epsilon_{abc} {\hat v}_c \nonumber\\ 
& +  \lambda \left[ f_E' \left( \kappa \right) - 3  \frac{f_E \left( \kappa \right)}{\kappa} \right] {\hat v}_e {\hat s}_e  
\; \left( D_i {\hat s} \right)^a \; \left( D_j {\hat s} \right)^b 
\; \epsilon_{abc} {\hat s}_c  \nonumber\\ 
&  +  \lambda \frac{f_E \left( \kappa \right)}{\kappa} 
\left\{ 
\left[ 
\left( D_i {\hat v} \right)^a \; \left( D_j {\hat s} \right)^b  + 
\left( D_i {\hat s} \right)^a \; \left( D_j  {\hat v} \right)^b 
\right]   
\epsilon_{abc}  {\hat s}_c +  
\left( D_i {\hat s} \right)^a \; \left( D_j {\hat s} \right)^b  \; \epsilon_{abc}  {\hat v}_c 
\right\} \;. 
\label{result-D-ground}
\end{align} 
In these expressions, we have used the combinations 
\begin{equation} 
\left(  D_i {\hat v} \right)^a \equiv  D_i^{ab} \, {\hat v}_b \;\;,\;\; 
\left( D_i {\hat v} {\hat s} \right) \equiv   D_i^{ab} \, {\hat v}_b \,  {\hat s}_a \;\;,\;\; 
\left( D_i   D_j  \right)^{ab} \equiv D_i^{ac} \, D_i^{cb} \;,\; \dots 
\end{equation} 

As we mentioned, these expressions are valid in the regime in which the product between the frequency and the arm lengths is much smaller than one, but do not assume that the product between the frequency and the separation distance between the two detectors, is also small (namely, $\kappa$ does not need to be $\ll 1$). When this is also true, our results simplify further into 
\begin{equation}
\lim_{\kappa \to 0} {\cal M}_{ij}^\lambda =  \frac{\left( D_i D_j \right)^{aa}}{5} \;\;\;,\;\;\; 
\lim_{\kappa \to 0} {\cal D}_{ij}^\lambda \left( {\hat v} \right) = 
\frac{2 \lambda \, \left( D_i D_j \right)^{ab} }{15} \epsilon_{abc}{\hat v}_c \,.
\label{small-f}
\end{equation} 

The analytic expressions (\ref{result-M-ground}) and (\ref{result-D-ground}) can be readily evaluated for any pair of detectors. In Table \ref{tab:detectors} in Appendix~\ref{app:detectors} we provide the explicit expressions for the vectors $\vec{s}_i ,\, {\hat U}_i ,\, {\hat V}_i$ for the two LIGO, the Virgo, and the KAGRA detectors. As an example, in Figure \ref{fig:LHLL-VK} we show the overlap functions for the pair of LIGO detectors (first row) and for the Virgo-KAGRA pair (second row).  The figure confirms the correctness of the covariant, analytical expressions  (\ref{result-M-ground}) and (\ref{result-D-ground}), 
obtained using both the analytical  expressions  given above and numerical evaluations. We have verified that the agreement between the analytic and numerical results persists for other generic directions of ${\hat v}$, beyond the particular choice in Fig.~\ref{fig:LHLL-VK}.
For the case of the monopole, 
equivalent formulas, but not covariant since they make use of a particular reference frame, can be found 
in \cite{Seto:2007tn,Seto:2008sr}. Our general results    identify clearly the `parity-violating' contributions 
proportional to the Levi-Civita tensor $\epsilon_{abc}$, and do not make any hypothesis on the 
shape of the detector (whose arms can form angles different than $90$ degrees).

\begin{figure}[t!]
\centering 
\includegraphics[width=0.45\textwidth]{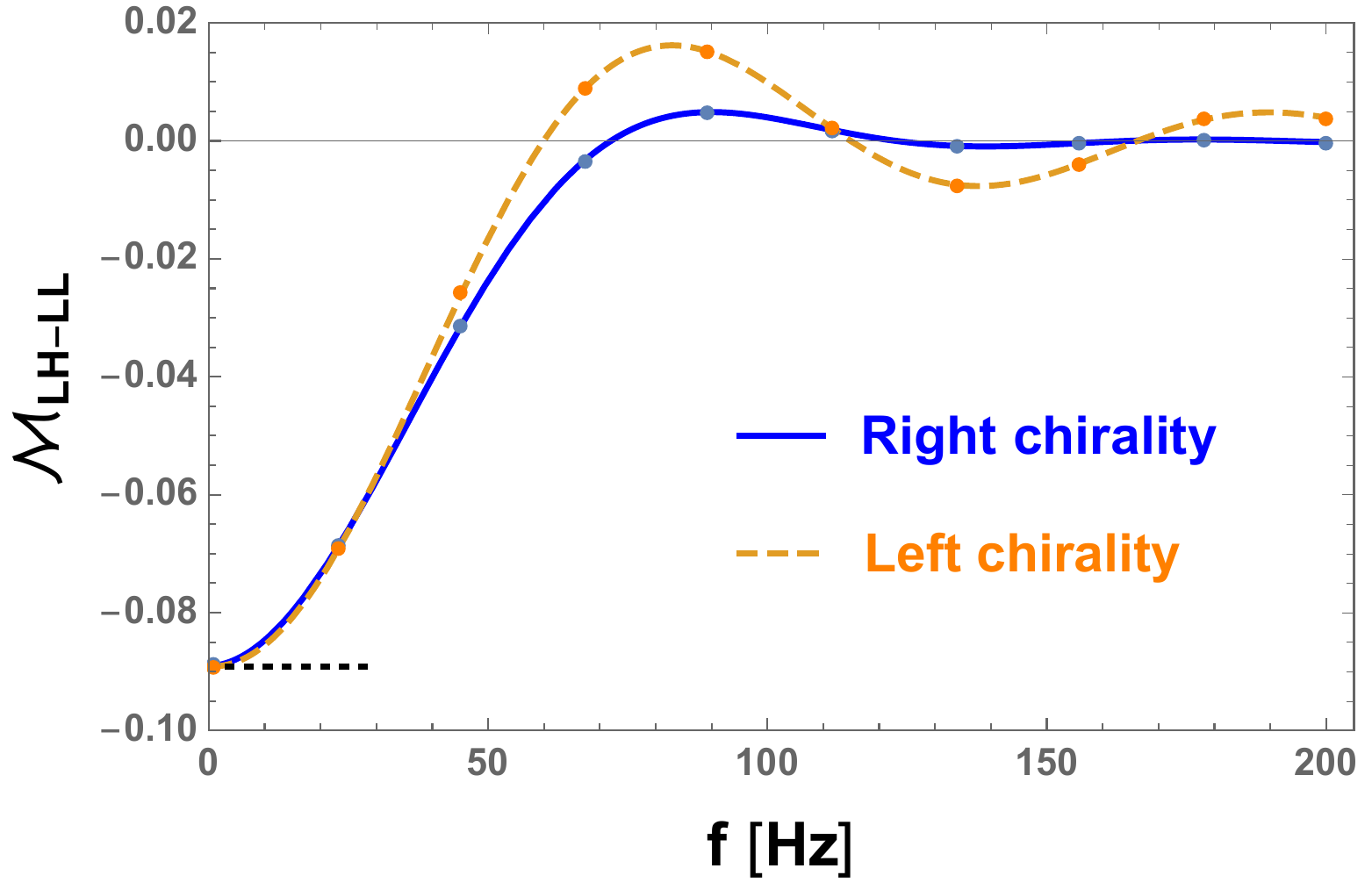}
\includegraphics[width=0.45\textwidth]{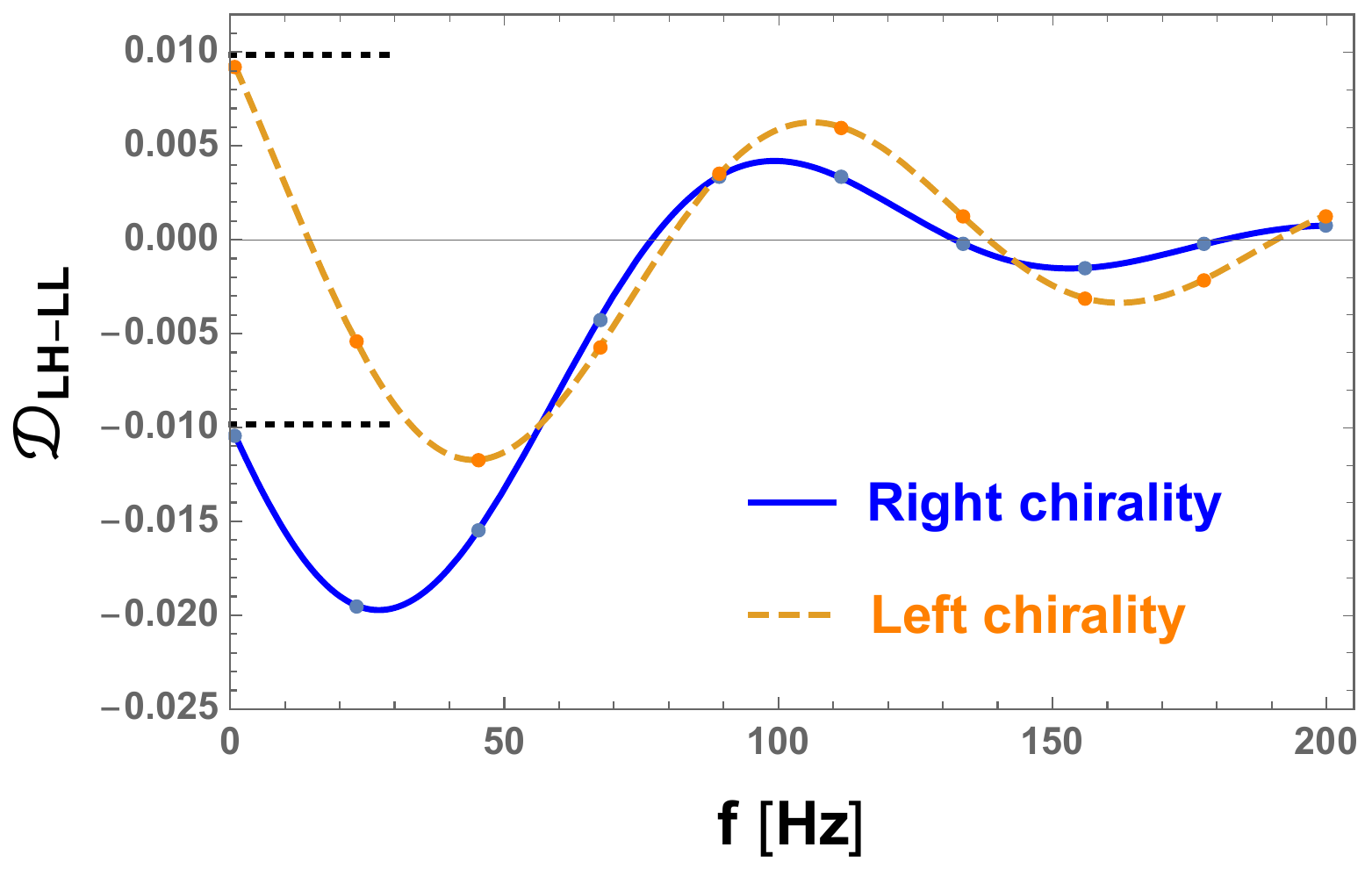} \\
\includegraphics[width=0.45\textwidth]{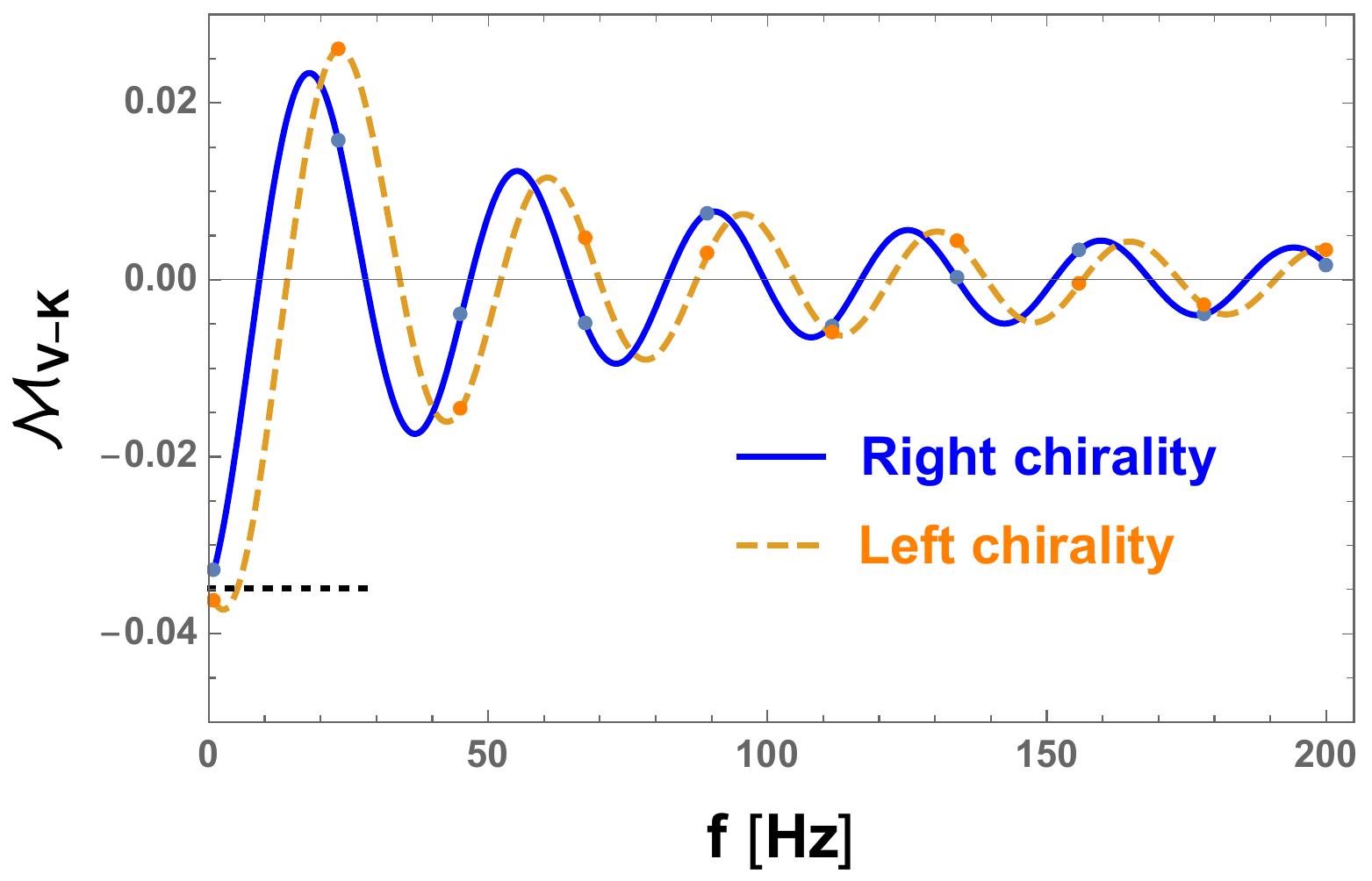}
\includegraphics[width=0.45\textwidth]{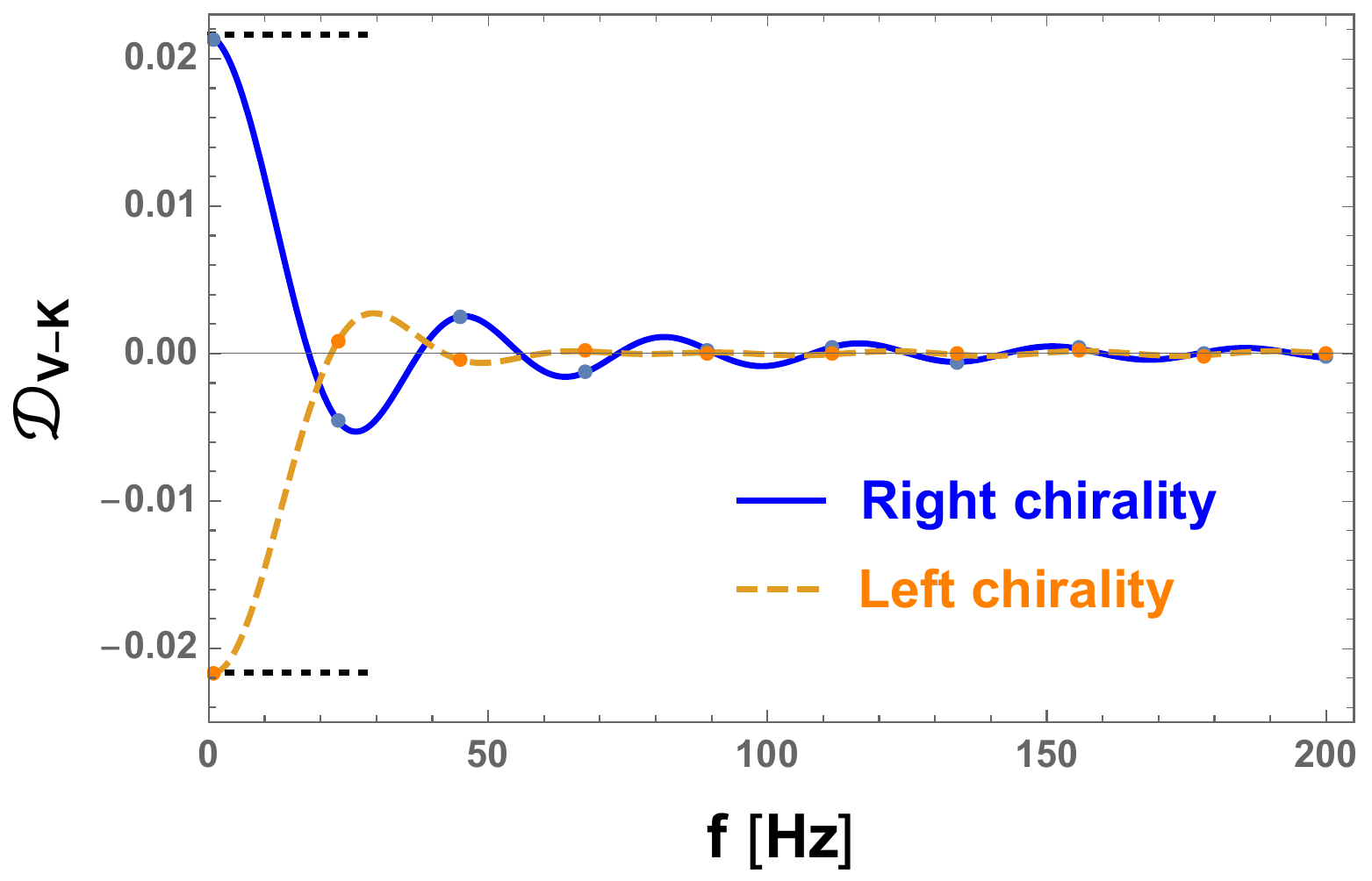} 
\caption{\it First row: monopole and dipole overlap functions for the LIGO Hanford (LH) and LIGO Livingston (LL) pair. 
Second row: monopole and dipole overlap functions for the Virgo (V) and KAGRA (K) pair. 
In the dipole case, ${\hat v} = \left( 0 ,\, 0 ,\, 1 \right)$ (in the coordinate system introduced in Appendix \ref{app:detectors}) has been chosen for illustrative purposes.  The solid lines  
are the analytic expressions (\ref{result-M-ground}) and  (\ref{result-D-ground}). The dotted black lines at small frequency are the asymptotic values (\ref{small-f}). The dots are obtained from a numerical evaluation. }
\label{fig:LHLL-VK}
\end{figure}

\subsection{Comments about the chiral contributions to the two-point overlap function~${\cal M}$}

The last two contributions to the monopole overlap function~(\ref{result-M-ground}), proportional to $f_D$ and $f_E$, distinguish between the two different GW polarizations and depend on the separation between interferometers as well as their orientations.  
{We note that they  vanish in the limit of coincident instruments (see  eq.~(\ref{small-f}))} or when the detector arms are oriented such that the quantity $D_i^{ac} \, D_j^{bd} $ is symmetric under the $a \leftrightarrow b$ exchange. The former condition can be easily understood: by measuring the GW at one location, one cannot determine how its profile changes as it propagates, and hence left- and right-handed GWs cannot be distinguished. A geometrical interpretation for the latter condition will be given below.

To obtain a  more explicit expression for the overlap function, we place and orient the detectors at the following coordinates (this choice can always be done  with no loss of generality),
\begin{align}
\hat x_1 & = \left( 1 ,\, 0 ,\, 0 \right), \nonumber \\ 
\hat U_1 & =  \left( 0 ,\, \sin \alpha ,\, \cos \alpha \right), \ \hskip2.5cm \quad 
\hat V_1 =  \left( 0 ,\, \cos \alpha ,\, - \sin \alpha \right) \,,  \nonumber \\
\hat x_2 & = \left( \cos \phi ,\, \sin \phi ,\, 0 \right),  \nonumber \\ 
{\hat U}_2 & =  \left( - \sin \phi \sin \beta ,\, \cos \phi \sin \beta ,\, \cos \beta \right), \quad 
{\hat V}_2 =  \left( - \sin \phi \cos \beta ,\, \cos \phi \cos \beta ,\, - \sin \beta \right) \,,  
\end{align}
where $0 \leq \phi \leq \pi$, and $0 \leq \alpha,\, \beta \leq 2 \pi$. The angles $\alpha$ and $\beta$ give the orientation of the ${\hat U}-$arm in terms of the angle from the north toward the east direction (where these directions are expressed at the location of each detector). With this choice, the unit vector going from the first to the second detector is 
\be
{\hat s} = \frac{1}{\sqrt{2 \left( 1 - \cos \phi \right)}} \; \left( -1 + \cos \phi ,\, \sin \phi ,\, 0 \right)\,,
\ee
and the $\lambda-$dependent terms in the monopole overlap function~(\ref{result-M-ground}) give rise to  
\begin{align}
\Delta \mathcal{M} & \equiv   \mathcal{M}_{ij}^+ -  \mathcal{M}_{ij}^- \nonumber \\
& = \frac{\kappa^2 \left( - 3 + \cos \phi \right) j_0 \left( \kappa \right) + \left[ 3 \left( 7 -\kappa^2 \right) + \left( 9 + \kappa^2 \right) \cos \phi \right] j_2 \left( \kappa \right)}{24 \kappa} \, \sin \left( \frac{\phi}{2} \right) 
\sin \left[ 2 \left( \alpha + \beta \right) \right] \,, 
\label{DRdif}
\end{align} 
where we note that $\phi$ is the angle (centered in the center of the Earth) between the two detectors, while $\alpha$ and $\beta$ express, respectively, the orientations of the $U-$arm of the two detectors.  {We notice that eq.~\eqref{DRdif} always vanishes when $\phi=0$ (the two detectors are coplanar) and when the sum $(\alpha+\beta)$ is  equal to zero or  $\pi/2$.  If this condition occurs, indeed, the combination  ${\cal D}_i^{ac} \, {\cal D}_j^{bd}$ is symmetric  in the indexes $(a,\,b)$: as we have discussed above, this implies null sensitivity to 
 parity violating effects. }
{This result can also interpreted geometrically as follows. If $\alpha=-\beta$, the system of detectors is symmetric about the plane through the maximal circle on Earth that passes halfway between the two detectors. As a consequence, a right-handed gravitational wave coming from one side of this plane is indistinguishable from a left-handed one coming from the opposite direction, so that the system, after selecting the isotropic monopole contribution, is insensitive to chirality. This argument is analogous to, and generalizes, that given in~\cite{Smith:2016jqs}, where it was shown that coplanar detectors are insensitive to chirality (in that case, the symmetry plane coincided with the plane of the two detectors).}

 In particular, if the detectors are located at the antipodes ($\phi = \pi$), the absolute value of eq.~\eqref{DRdif} is maximized and reduces to
\begin{equation}
\Delta  \mathcal{M}_{\rm antipodes} =  \frac{-\kappa^2 \, j_0 \left( \kappa \right) + \left( 3 - \kappa^2 \right) j_2 \left( \kappa \right)}{6 \kappa} \sin \left[ 2 \left( \alpha + \beta \right) \right] \;.  \label{DRanti}
\end{equation} 
In what comes next, using our formulas we discuss more quantitatively the best choices of location for antipodal ground based detectors in order to detect parity violating effects in the SGWB. Similar considerations can also be found
in \cite{Seto:2007tn,Seto:2008sr}.
\begin{figure}[t!]
\begin{center}
\includegraphics[width = 0.5 \textwidth]{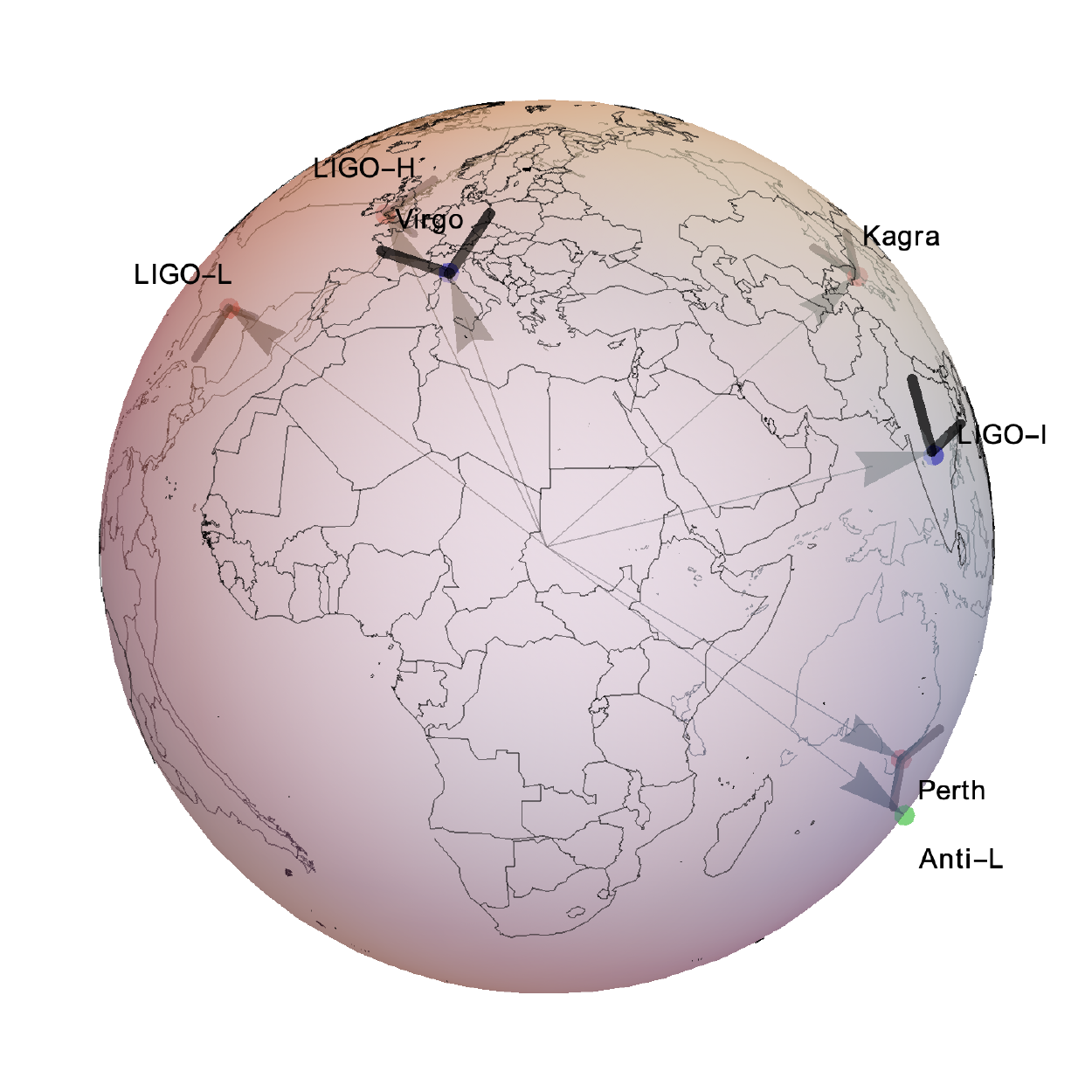}
\caption{\it 
The location of all {existing} detectors on Earth, together with a LIGO-India detector in Maharashtra, and a hypothetical optimal-for-chiral-SGWB detector in Perth. We also show the antipodes of the LIGO-Livingston detector (green dot), which is not far from the Perth detector.  We note that the Figure shows the point of view of an observer at a specific location in space, who sees less than half of the Earth. Lighter lines  {(red dots)} are used to indicate continents {(interferometers)} that are not seen by this observer. 
}
\label{fig:EarthSphere}
\end{center} 
\end{figure}

\subsubsection*{Choice of Earth location for optimal detection of a chiral SGWB}

If we search for the antipodes of the four known detectors (Hanford, Livingston, Virgo, KAGRA), we see that all of them fall in the Ocean (Pacific, Atlantic and Indian). The antipode of LIGO-Livingston (L) falls in the Indian Ocean near Australia. The closest large city to it is Perth (P). Let us compute the optimal overlap function for this pair of detectors. Recall that, in our coordinate system, defined in App.~\ref{app:detectors}, LIGO-Livingston (L) is located at 
\begin{equation}
\vec{x}_{\rm L} = R \left( -0.011 ,\, -0.860 ,\, 0.508 \right) \,,
\end{equation}
{with $R$ denoting the radius of the earth,} and its arms are directed along 
\begin{equation}
{\hat u}_{\rm L} = \left( -0.953 ,\, -0.144 ,\, -0.266 \right) \;\;,\;\; 
{\hat v}_{\rm L} = \left( 0.302 ,\, -0.488 ,\, -0.819 \right)\,.
\end{equation} 
Moreover, in our coordinate system, P is located at 
\begin{equation}
\vec{x}_{\rm P} = R \left( -0.370 ,\, 0.763 ,\, -0.529 \right) \,,
\end{equation}
which gives a distance 
\begin{equation}
s = \vert \vec{x}_p - \vec{x}_{L} \vert \simeq 1.96 \, R \;\; \Rightarrow \;\; \kappa \simeq  \frac{f }{24\,\rm Hz} \,.
\end{equation}
Therefore the two detectors are nearly opposite, as can be seen in Fig.~\ref{fig:EarthSphere}. 

We now place the arm ${\hat u}_P$  at the angle $\alpha$ from the north direction towards east (from the point of view of an observer at P), while  ${\hat v}_P$ is at the angle $\frac{\pi}{2} + \alpha$. We then have 
\begin{eqnarray}
{\hat u}_{\rm P} &=& \cos \alpha \left(-0.230, 0.476, 0.848 \right) 
+ \sin \alpha  \left(-0.899, -0.436, 0 \right) \,, \nonumber\\ 
{\hat v}_{\rm P} &=& - \sin \alpha \left(-0.230, 0.476, 0.849 \right) 
+ \cos \alpha  \left(-0.899, -0.436, 0 \right)  \,. 
\end{eqnarray} 
The difference in the overlap function $\Delta  \mathcal{M}$ for the L-P pair gives 
\begin{eqnarray} 
\Delta  \mathcal{M} &=& 
\frac{
\kappa \left( 1 - 0.31 \kappa^2 \right) \cos \kappa
+  \left( - 1 + 0.64 \kappa^2 \right) \sin \kappa  }{\kappa^4} 
\left[ -0.22 \cos \left( 2 \alpha \right) + 1.5 \sin \left( 2 \alpha \right) \right] \,.\nonumber\\ 
\end{eqnarray} 
\begin{figure}[t!]
\begin{center}
\includegraphics[width = 0.48 \textwidth]{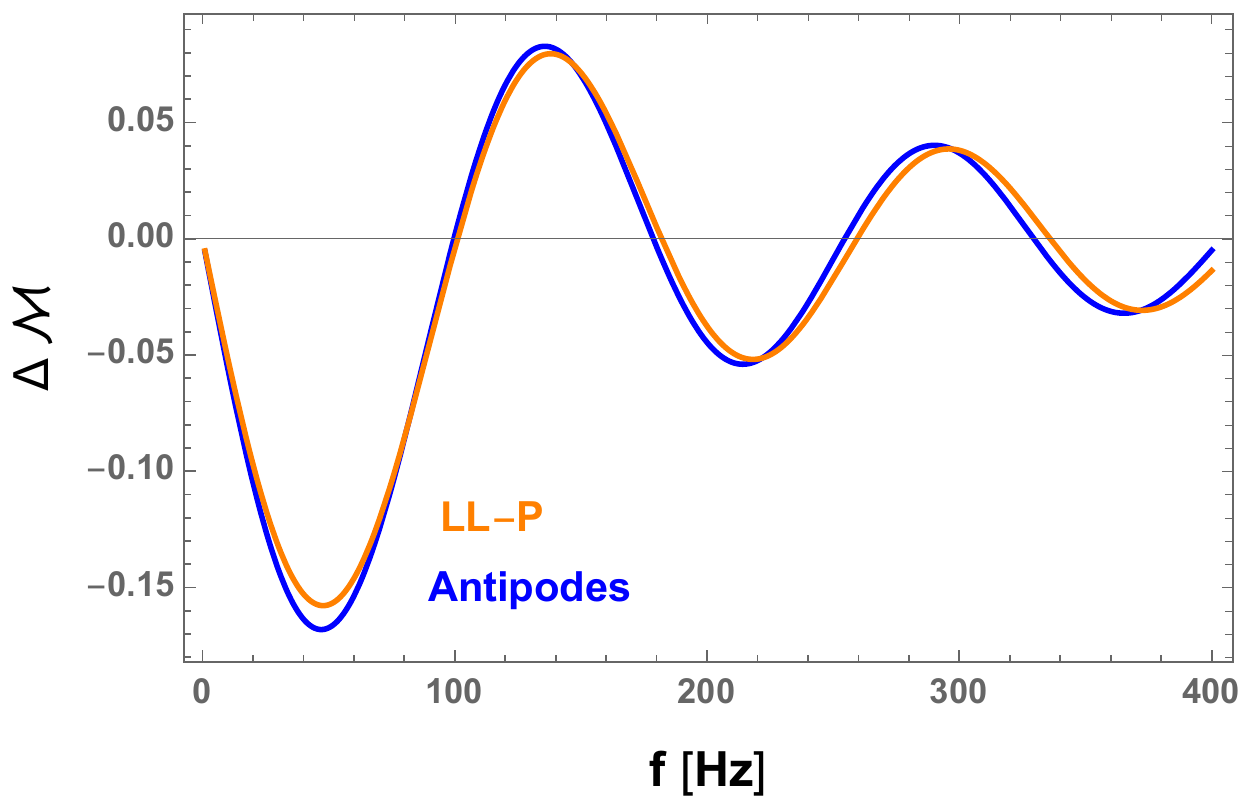}
\caption{\it  The function  {$\Delta {\cal M}$}, 
sensitive to parity violation  (difference of the overlap functions of opposite chirality, see eq.~\eqref{DRanti}) of two ideal detectors at the antipodes, and of LIGO-Livingston with a detector at Perth, Australia. By expanding the $\kappa$ dependent part of eq.~\eqref{DRanti} for large $\kappa$, we find that the zeros of this function occur at the frequencies $f \simeq \frac{\pi}{d} \left( \frac{1}{2} + n \right)$, where $d$ is the diameter of the Earth and $n$ is an integer number. By comparing with the figure, one can see that this relation works well already at $n=1$. }
\label{fig:antipodes-LLP}
\end{center} 
\end{figure}
We then have for:
\begin{eqnarray}
& & {\rm antipodes} \;,\; \alpha_{\rm best} = \frac{\pi}{4} \;\; \Rightarrow \;\; \Delta {\cal M} = 1.5 \,\frac{ \kappa \left( - 1 +  \kappa^2/3 \right) \cos \kappa + \left( 1 - 2 \kappa^2/3 \right) \sin \kappa}{\kappa^4} \nonumber\\ 
& & \quad\quad  \quad\quad  \quad
{\rm with } \;  \kappa = \frac{f}{23.5\,\rm Hz} \,, \nonumber\\ 
& & {\rm L-P} \;,\; \alpha_{\rm best} = 2.43 \;\; \Rightarrow \;\; 
\Delta {\cal M} = 1.56 \,\frac{
\kappa \left( -1 + 0.31 \kappa^2 \right) \cos \kappa 
+  \left(  1 - 0.64  \kappa^2 \right) \sin \kappa }{\kappa^4}  \nonumber\\
& & \quad\quad  \quad\quad  \quad
{\rm with } \;  \kappa =  \frac{f}{24\,\rm Hz}  \,.
\end{eqnarray} 
At small frequencies $ \kappa \ll 1$ this yields
\begin{equation}
\Delta  \mathcal{M}_{\rm antipodes}{(\alpha_\text{best})} \simeq - \frac{f}{177\,\rm Hz} \;\;,\;\; 
\Delta  \mathcal{M}_{\rm L-P}{(\alpha_\text{best})}\simeq -  \frac{f}{191\,\rm Hz} \,.
\label{eq:DR-optimal}
\end{equation} 
Consequently, an additional GW detector close to Perth, Australia, rotated clockwise by $2.43$ radiants from the local north direction, is essentially an optimal choice to measure parity with a network of ground-based detectors.

The expressions~\eqref{DRdif} and \eqref{eq:DR-optimal} can be employed to determine the SNR of detecting a net polarization in the SGWB. The difference in the frequency dependence of the response functions ${\cal M}^+$ and ${\cal M}^-$ can be utilized to distinguish a chiral from a non-chiral SGWB. This analysis (using numerically evaluated response functions)  was performed in Ref.~\cite{Crowder:2012ik} for the Hanford and Livingston LIGO, VIRGO and KAGRA detectors and for a power-law signal. In the specific case of a frequency-independent SGWB, it was found that   maximal chirality can be detected or excluded for an amplitude up to $\Omega_{GW} \gtrsim 10^{-8}$. It would be interesting to extend this analysis to include an antipodal detector with the optimal orientation $\alpha_\text{best}$, but this is beyond the scope of the present paper.

\subsection{SNR for the Einstein Telescope}
\label{sec:ET}

The  Einstein Telescope is a proposal for a ground-based interferometer with a triangular shape 
 with arm length  $L=10\,$km. It  will be an observatory of the third generation aiming to reach a sensitivity for GW signals emitted by astrophysical and cosmological sources about a factor of ten better than the currently operating ground based detectors. It will be formed by three detectors, each in turn composed of two interferometers (xylophone configuration)~\cite{Punturo:2010zz, Sathyaprakash:2012jk}.
The triangular planar configuration of ET, similar to LISA, allows to use the same approach developed   in Section ~\ref{sec: sec_LISA}  to compute the SNR for measuring the circular polarization. 
 For the computation we use eq.~\eqref{snr_lisa}, where we consider  the noise power spectrum $ P_n^{\rm ET}(f)$  for a third-generation gravitational wave interferometer~\cite{Hild_2011}. The expression for the SNR, for a scale invariant $\Omega_{GW}^\lambda$, in this case is
\begin{eqnarray}
{\rm {SNR}}_\text{ET}&=&\,\frac{3H_0^2}{2\pi^2} \, v\,\left|\sum_\lambda\lambda\,\Omega_{GW}^\lambda\right|\,\left[2\int dT\,\cos^2\alpha(T)\int_0^\infty\frac{df}{f^4}\frac{D(fL)^2}{(f\,P_n^{\rm ET}(f))^2}\right]^{1/2}\nonumber\\
&\approx& 7.5\times 10^{13}\,v\left|\sum_\lambda\lambda\,\Omega_{GW}^\lambda h^2\right|\,\left[\int_0^{\frac{T}{1\,{\rm {year}}}}\cos^2\alpha(x)\,dx\right]^{1/2} \,.
\end{eqnarray}
where for the dipole response function $D(fL)$ we have used the value at small frequency given by eq.~\eqref{eq:small_frequency}.

 Comparing the sensitivity to circular polarization for operating ground-based detectors, derived in Ref.~\cite{Crowder:2012ik}, with ET, we note that the improved sensitivity of the Einstein Telescope, in particular at low frequencies, enables to out-perform the current LIGO configuration, taking into account the expected magnitude of the kinematic dipole of $v \sim 10^{-3}$. This nicely demonstrates the important interplay between detector sensitivity, location and co-planarity for ground-based detectors. With two copies of the Einstein Telescope (or of the Cosmic Explorer~\cite{Reitze:2019iox}), one could of course benefit from increased sensitivity and the elimination of the dipole factor $v$ since the monopole is already sensitive to chirality. 

%

\section{Conclusions}
\label{sec: conclusions}

The detection of the SGWB is a major goal for GW interferometers, which is expected to be achieved in the coming years. On the other hand, the amplitude and properties of the cosmological SGWB are highly model dependent.  Any detection or constraint on this cosmological SGWB will contain valuable information about the early Universe. In this situation, it is crucial to extract and characterize all properties of any SGWB detected. In this paper we focus on the ability of ground- and space-based detectors to measure the net polarization of the SGWB, which could be a smoking gun for parity violating interactions in the early Universe.

For an isotropic SGWB, a system of coplanar detectors is insensitive to the polarization of the SGWB~\cite{Seto:2007tn,Seto:2008sr,Smith:2016jqs}. Making the symmetries of the response functions of ground- and space-based detectors explicit, we provide a transparent demonstration of this result as well as of the two possibilities to circumvent it: (i) for planar detectors (such as LISA or ET), we make use of the kinematic dipolar anisotropy induced by the motion of the solar system with respect to the cosmic rest frame~\cite{Allen:1996gp,Kudoh:2005as,Seto:2006hf} and (ii) for a network of ground-based detector, the curvature of the earth breaks co-planarity~\cite{Seto:2007tn,Seto:2008sr,Crowder:2012ik}. In the present work we reconsider previous results by taking into account the full response functions and noise curves in the entire frequency band (for planar detectors). Moreover, we  provide fully analytical and covariant expressions for the (parity-sensitive) response functions of a ground-based detector network. 

We find that LISA and ET, despite operating at very different frequencies, will have a similar sensitivity to a scale-invariant SGWB, and could detect an ${\cal O}(1)$ net polarization in a SGWB with a magnitude of $\Omega_{GW} h^2 \simeq 10^{-11}$ with an SNR of order one. We emphasize that these two instruments should be seen as complementary probes, since 
the SGWB may vary significantly between the LISA and ET frequency bands. For both LISA and ET, the auto-correlation channels are blind to chirality and the entire sensitivity stems from cross-correlating the two TDI channels. 

For a network of ground-based detectors we provide fully covariant analytical expressions for the monopole and dipole response functions. It is much more rapid to evaluate these analytic expressions than to compute numerically the angular integrals that are needed to obtain the response functions numerically, and therefore we hope that these analytic relations can be used to speed up future studies of the SGWB polarization. Since the sensitivity to net polarization of the (dominant) monopole contribution to the SGWB arises from the departure from co-planarity, the detector location and orientation plays a crucial role. 

In summary, in this paper we studied a specific feature that can contribute to the characterization of the SGWB: the possibility for measuring a circular polarization degree of a gravitational wave background through a dipolar modulation induced by the motion of the reference frame with respect to the cosmic frame. This could help single out specific cosmological mechanisms characterized by violation of parity in the early universe, and it is therefore an interesting observational target for current and future interferometers. As future work, it would be interesting to apply this approach to other  GW detectors, which can be sensitive to the circular
polarization using this method. For example,  the proposed Japanese space-based GW observatory DECIGO \cite{Kawamura:2011zz}, or   
 to astrometric GW observations  
that aim to reveal effects induced by a SGWB using data from the Gaia mission \cite{Mihaylov:2018uqm}.

\vspace{1cm}

\subsection*{Acknowledgments}

We warmly thank the LISA Cosmology Research Group for many stimulating and insightful discussions on various aspects of the SGWB.
V.D.\ acknowledges funding by the Deutsche Forschungsgemeinschaft under Germany's Excellence Strategy - EXC 2121 ``Quantum Universe'' - 390833306.  JGB acknowledges support from the Research Project FPA2015-68048-03-3P(MINECO-FEDER) and the Centro de Excelencia Severo Ochoa Program SEV-2016-0597. M.Pi. acknowledges the support of the Spanish MINECOs ``Centro de Excelencia Severo Ochoa'' Programme under grant SEV-2016-059.
This project has received funding from the European Unions Horizon 2020 research and innovation programme under the Marie Sk\l{}odowska-Curie grant agreement No 713366. The work of L.S. is partially supported by the US-NSF grant PHY-1820675. G.T. is partially supported by STFC grant ST/P00055X/1.

\bigskip
\begin{appendix}

\setcounter{equation}{0}
\renewcommand{\theequation}{\thesection\arabic{equation}}

\section{GW polarization operators in the chiral basis} 
\label{app:GW-polarization-operators} 

We follow the standard definition of the GW polarization operators, that we summarized in the 
  work \cite{Bartolo:2018qqn}. It is straightforward to show that the opertors can be also introduced as 
\begin{eqnarray}
e_{ab,\lambda} \left( {\hat k} \right) = e_{a,\lambda} \left( {\hat k} \right) \,  e_{b,\lambda} \left( {\hat k} \right) \equiv 
\frac{ {\hat u}_a \left( {\hat k} \right) + i  \lambda \, \,  {\hat v}_a \left( {\hat k} \right) }{\sqrt{2}} \, 
\frac{ {\hat u}_b  \left( {\hat k} \right) + i  \lambda \,  {\hat v}_b  \left( {\hat k} \right)}{\sqrt{2}} \,, \quad 
\label{u-v-oneindex} 
\end{eqnarray} 
where we recall that $\lambda = + 1$ (respectively, $\lambda = - 1$) correspond to the right-handed (respectively, the left-handed) helicity,~\footnote{Another basis that is often chosen for the polarization operators is the  $\{+,\times\}$ basis, related to the chiral basis by $e_{ab,\lambda} = \frac{e_{ab}^{(+)} + \lambda i e_{ab}^{(\times)}}{\sqrt{2}}$.}  and where 
\begin{equation}
{\hat u} \left( {\hat k} \right) \equiv \frac{ {\hat k} \times {\hat e}_z  }{\vert  {\hat k} \times {\hat e}_z \vert}  \;\;\;\;,\;\;\;\; 
{\hat v} \left( {\hat k} \right) \equiv {\hat k} \times {\hat u} \left( {\hat k} \right) =   \frac{ \left( {\hat k} \cdot {\hat e}_z \right) \, {\hat k} - {\hat e}_z }{\vert  {\hat k} \times {\hat e}_z \vert}  \;, 
\label{uv-choice} 
\end{equation} 
where ${\hat e}_z$ is the unit vector along the third-axis. 

It immediately follows that 
\begin{eqnarray}
e_{ab,\lambda}^* (  {\hat k} ) = e_{ab,\lambda} ( - {\hat k} ) = e_{ab,-\lambda} (  {\hat k} )  
\;\; \;\;  ,\;\; \;\;  e_{ab,\lambda}^* ( {\hat k} )\, e_{ab,\lambda'} ( {\hat k} ) = \delta_{\lambda \lambda'} \;. 
\label{e-properties}
\end{eqnarray} 

Moreover, one can verify by direct inspection that 
\begin{align}
e_{i,\lambda}(\hat{k})e_{i',\lambda}(-\hat{k})=-\frac{1}{2}\left(\delta_{ii'}-\hat{k}_i\,\hat{k}_{i'}-i\lambda\,\epsilon_{ii'j}\hat{k}_j\right)\,.
\label{e-1index-identity}
\end{align}

Combining this identity with eq.~(\ref{u-v-oneindex}), we can also write 
\begin{align} 
e_{ab,\lambda} \left( {\hat k } \right)  \,  e_{cd,\lambda} \left( - {\hat k } \right)   = & \frac{1}{4} 
\left[ \delta_{ac} - {\hat k}_a {\hat k}_c - i \lambda \epsilon_{ace} {\hat k}_e \right] 
\left[ \delta_{bd} - {\hat k}_b {\hat k}_d - i \lambda \epsilon_{bdf} {\hat k}_f \right] \nonumber\\ 
 = &  \frac{1}{4} \left[ 
\left( \delta_{ac} - {\hat k}_a {\hat k}_c \right)  \left( \delta_{bd} - {\hat k}_b {\hat k}_d \right) 
+ \left( \delta_{ad} - {\hat k}_a {\hat k}_d \right)  \left( \delta_{bc} - {\hat k}_b {\hat k}_c \right) \right. \nonumber \\
& \left. - \left( \delta_{ab} - {\hat k}_a {\hat k}_b \right)  \left( \delta_{cd} - {\hat k}_c {\hat k}_d \right) \right] \nonumber\\ 
& - \frac{i \lambda}{4} \left[ 
\left( \delta_{ac} - {\hat k}_a {\hat k}_c \right) \epsilon_{bdf} {\hat k}_f 
+ \left( \delta_{bd} - {\hat k}_b {\hat k}_d \right) \epsilon_{ace} {\hat k}_e \right] \,. 
\label{e-2indices-identity}
\end{align}

\section{Comparison with previous computation}
\label{app:comparison}
\setcounter{equation}{0}

This Appendix  provides a detailed comparison betweeen our results  of Section \ref{SNR_LISA}  and the findings of Seto in  \cite{Seto:2006hf} for the magnitude of the signal-to-noise ratio associated with measurements of the SGWB   circular polarization with LISA.  The comparison is made difficult by the different notation used in the two works.   Our aim is to carry on all the steps that
allow us to re-write the results of  \cite{Seto:2006hf}  using the notation implemented in our paper. Our conclusion will be that 
our findings for the magnitude of the signal-to-noise ratio is a factor of 10 larger than \cite{Seto:2006hf}.
 We use a superscript ${}^{(S)}$ to denote quantities in Seto's work  \cite{Seto:2006hf}.
 
We start from our decomposition for tensor fluctuations,  
\begin{align}
h_{ij}(\vec{x},t)=\int d^3 k e^{-2\pi i \vec{k}\cdot\vec{x}}\sum_P e_{ij,P}(\hat k)\left[e^{2\pi i k t}h_P(k,\hat k)+e^{-2\pi i k t}h_P^*(k,-\hat k)\right]\,,
\end{align}
where we decompose in $P=+,\,\times$ polarizations instead of $L,\,R$. In \cite{Seto:2006hf}  the notation is 
\begin{align}
h_{ij}(\vec{x},t)=\int_{-\infty}^\infty df\,d^2 \hat n\, e^{2\pi i f\left(t- \hat\bn\cdot\bx\right)}\sum_P {}^{(S)}e_{ij,P}(\hat\bn){}^{(S)}h_P(f,\hat\bn)\,,
\end{align}
where ${}^{(S)}e_{ij,P}(\hat{n})=\sqrt{2}\,e_{ij,P}(\hat{n})$.

To proceed with our comparison, we separate our expressions for $h_{ij}$  into (we use $\hat{n}=\hat{k}$)
\begin{align}
h_{ij}(\vec{x},t)& =\int_0^\infty k^2\,dk\,d^2 \hat n e^{-2\pi i k\hat{n}\cdot\vec{x}}\sum_P e_{ij,P}(\hat{n})e^{2\pi i k t}h_P(k,\hat{k})
\nonumber
\\
&
+\int_0^\infty k^2\,dk\,d^2 \hat n e^{-2\pi i k\hat{n}\cdot\vec{x}}\sum_P e_{ij,P}(\hat{n})e^{-2\pi i k t}h_P^*(k,-\hat{k})\,, \nonumber\\
&=\int_0^\infty f^2\,df\,d^2 \hat n e^{-2\pi i f\hat{n}\cdot\vec{x}}\sum_P e_{ij,P}(\hat{n})e^{2\pi i f t}h_P(f,\hat{n})\nonumber\\
&+\int^0_{-\infty} f^2\,df\,d^2 \hat n e^{2\pi i f\hat{n}\cdot\vec{x}}\sum_P e_{ij,P}(\hat{n})e^{2\pi i f t}h_P^*(-f,-\hat{n})
\,,
\nonumber\\
&=\int_0^\infty f^2\,df\,d^2 \hat n e^{-2\pi i f\hat{n}\cdot\vec{x}}\sum_P e_{ij,P}(\hat{n})e^{2\pi i f t}h_P(f,\hat{n})\nonumber\\
&+\int^0_{-\infty} f^2\,df\,d^2 \hat n e^{-2\pi i f\hat{n}\cdot\vec{x}}\sum_P e_{ij,P}(-\hat{n})e^{2\pi i f t}h_P^*(-f,\hat{n})\,,
\end{align}
where in the last step we have changed $\hat{n}\to-\hat{n}$ in the second integral. 

To compare with \cite{Seto:2006hf},  we can make  the identification
\begin{align}
\sqrt{2}\,\,{}^{(S)}h_P(f,\hat{n})=f^2\left\{
\begin{array}{ll}
h_P(f,\hat{n}) & f>0 \\
(-1)^P h_P^*(-f,\hat{n}) & f<0
\end{array}\right.
\end{align}
where $(-1)^P$ is $+1$ for $P=+$ and $-1$ for $P=\times$ .
 In order to  prove this fact, we follow \cite{Seto:2006hf}, and write
\begin{align}
\hat{n}=(\sin\theta\,\cos\phi,\,\sin\theta\,\sin\phi,\,\cos\theta)\,,
\end{align}
and
\begin{align}
&{e}_\theta=\partial_\theta\hat{n}=(\cos\theta\,\cos\phi,\,\cos\theta\,\sin\phi,\,-\sin\theta)\,, \nonumber\\
&{ e}_\phi=\partial_\phi\hat{n}=(-\sin\theta\,\sin\phi,\,\sin\theta\,\cos\phi,\,0)\,.
\end{align}
We have the relations 
${ e}^+\,=\,{ e}_\theta \,{ e}_\theta-{ e}_\phi\,{ e}_\phi$, ${ e}^\times\,=\,{ e}_\theta \, { e}_\phi+{ e}_\phi\,{ e}_\theta$. On the other hand,  $\hat{n}\to -\hat{n}$ is equivalent to $\theta\to\pi-\theta$, $\phi\to \phi+\pi$. This means that
\begin{align}
&{e}_\theta\to ((-\cos\theta)\,(-\cos\phi),\,(-\cos\theta)\,(-\sin\phi),\,-\sin\theta)={e}_\theta \nonumber \, ,\\
&{ e}_\phi\to(-\sin\theta\,(-\sin\phi),\,\sin\theta\,(-\cos\phi),\,0)=-{ e}_\phi\,,
\end{align}
which then implies ${e}^+\to { e}^+$, but ${e}^\times \to-{ e}^\times$. 
The work \cite{Seto:2006hf} defines
\begin{align}
\frac{i}{2}\langle {}^{(S)}h_+(f,\hat{n}){}^{(S)}h_\times^*(f',\hat{n}') -{}^{(S)}h_+^*(f,\hat{n}){}^{(S)}h_\times(f',\hat{n}') \rangle=\delta(f-f')\,\frac{\delta(\hat{n}-\hat{n}')}{4\pi}V(f,\hat{n})\,.
\end{align}
We now translate this expression in our notation
\begin{align}\label{ourcorrinV}
&{i}\langle h_+(f,\hat{n}){}h_\times^*(f',\hat{n}') -h_+^*(f,\hat{n})h_\times(f',\hat{n}') \rangle=2\frac{\delta(f-f')}{f^4}\,\frac{\delta(\hat{n}-\hat{n}')}{2\pi}V(f,\hat{n})\,\qquad f>0\,,\nonumber\\
&i\langle h_+^*(-f,\hat{n}){}h_\times(-f',\hat{n}') -h_+(-f,\hat{n})h_\times^*(-f',\hat{n}') \rangle=-2\frac{\delta(f-f')}{f^4}\,\frac{\delta(\hat{n}-\hat{n}')}{2\pi}V(f,\hat{n})\,\qquad f<0\,,
\end{align}
where these two expressions are actually one the complex conjugate of the other. 

\smallskip

We now proceed to compute expressions in terms $h^{L,R}$ modes. We have
\begin{align}
h^+=\frac{1}{\sqrt{2}}\left(h_R+h_L\right)\,,\qquad h^\times=\frac{i}{\sqrt{2}}\left(h_R-h_L\right)\,,
\end{align}
so that
\begin{align}
&i\left(h_+{}h_\times^* -h_+^*h_\times \right)=\left|h_R\right|^2-\left|h_L\right|^2\,.
\end{align}

The two point function found in eq.~(\ref{eq:h2pt}) reads 
\begin{eqnarray}
&& \langle h^\sigma(\vec{k},\,\tau)\,h^{\sigma'}(\vec{k}',\,\tau')\rangle=\delta_{\sigma\sigma'}\frac{\delta^{(3)}(\vec{k}+\vec{k}')}{4\pi\,k^3}\nonumber\\
&&\qquad\qquad\qquad\times \left\{P^\sigma(k)\,\cos[2\pi k(\tau-\tau')]-i(\hat{k}\cdot \vec{v})\,\left[2P^\sigma(k)-k\,P^\sigma{}'(k)\right]\,\sin[2\pi k(\tau-\tau')]\right\}\nonumber\,,\\ \label{apcoP}
\end{eqnarray}
with
\begin{align}
h^\sigma(\vec{k},\,\tau)=e^{2\pi i k t}h_\sigma(k,\hat{k})+e^{-2\pi i k t}h_\sigma^*(k,-\hat{k})\,,
\end{align}
Then the LHS of equation \eqref{apcoP}  rewrites
\begin{align}
&\langle \left[e^{2\pi i k \tau}h_\sigma(k,\hat{k})+e^{-2\pi i k \tau}h_\sigma^*(k,-\hat{k})\right]\left[e^{2\pi i k' \tau'}h_\sigma(k',\hat{k}')+e^{-2\pi i k' \tau'}h_\sigma^*(k',-\hat{k}')\right]\rangle\nonumber\\
&=\langle e^{2\pi i (k \tau+k'\tau')}h_\sigma(k,\hat{k})h_\sigma(k',\hat{k}')+e^{-2\pi i (k \tau+k'\tau')}h_\sigma^*(k,-\hat{k})h_\sigma^*(k',-\hat{k}')\nonumber\\
&+e^{2\pi i (k \tau-k'\tau')}h_\sigma(k,\hat{k})h_\sigma^*(k',-\hat{k}')+e^{-2\pi i (k \tau-k'\tau')}h_\sigma^*(k,-\hat{k})h_\sigma(k',\hat{k}')\rangle\,.
\end{align}
Now we note that this quantity must be a linear combination of $\sin[2\pi k (\tau-\tau')]$ and $\cos[2\pi k (\tau-\tau')]$. 
 This implies that
\begin{align}
\langle h_\sigma(k,\hat{k})h_\sigma(k',\hat{k}')\rangle=\langle h_\sigma^*(k,-\hat{k})h_\sigma^*(k',-\hat{k}')\rangle\,=\,0\,,
\end{align}
since these terms multiply cosines and sines of $2\pi k(\tau+\tau')$, and
\begin{align}
&\langle \cos[2\pi i (k \tau-k'\tau')]\Big[h_\sigma(k,\hat{k})h_\sigma^*(k',-\hat{k}')+h_\sigma^*(k,-\hat{k})h_\sigma(k',\hat{k}')\Big]\nonumber\\
&+i\sin[2\pi i (k \tau-k'\tau')]\Big[h_\sigma(k,\hat{k})h_\sigma^*(k',-\hat{k}')-h_\sigma^*(k,-\hat{k})h_\sigma(k',\hat{k}')\Big]\rangle\nonumber\\
&=\frac{\delta^{(3)}(\vec{k}+\vec{k}')}{4\pi\,k^3}\,\left\{P^\sigma(k)\,\cos[2\pi k(\tau-\tau')]-i(\hat{k}\cdot \bv)\,\left[2P^\sigma(k)-k\,P^\sigma{}'(k)\right]\,\sin[2\pi k(\tau-\tau')]\right\}
\end{align}
where we note in passing that the quantity in the second square bracket does not vanish.

By comparing the time dependent parts we thus get the time-independent correlators
\begin{align}
&\langle h_\sigma(k,\hat{k})h_\sigma^*(k',-\hat{k}')+h_\sigma^*(k,-\hat{k})h_\sigma(k',\hat{k}')\rangle=\frac{\delta^{(3)}(\vec{k}+\vec{k}')}{4\pi\,k^3}\,P^\sigma(k)\nonumber\\
&\langle h_\sigma(k,\hat{k})h_\sigma^*(k',-\hat{k}')-h_\sigma^*(k,-\hat{k})h_\sigma(k',\hat{k}')\rangle=-\frac{\delta^{(3)}(\vec{k}+\vec{k}')}{4\pi\,k^3}\,(\hat{k}\cdot \vec{v})\,\left[2P^\sigma(k)-k\,P^\sigma{}'(k)\right]\,.
\end{align}
We can  then conclude that
\begin{align}\label{finalhkhk}
\langle h_\sigma(k,\hat{k})h_\sigma^*(k',\hat{k}')\rangle=\frac{\delta^{(3)}(\vec{k}-\vec{k}')}{8\pi\,k^3}\,\Big\{P^\sigma(k)-(\hat{k}\cdot \vec{v})\,\left[2P^\sigma(k)-k\,P^\sigma{}'(k)\right]\Big\}\,.
\end{align}

\smallskip

Let now compute $\langle |h_R|^2-|h_L|^2\rangle$ as
\begin{align}
\langle h_R(k,\hat{k})h_R^*(k',\hat{k}')-h_L(k,\hat{k})h_L^*(k',\hat{k}')\rangle=\frac{\delta^{(3)}(\vec{k}-\vec{k}')}{8\pi\,k^3}\,\Big\{\Delta P^\sigma(k)-(\hat{k}\cdot \vec{v})\,\left[2\Delta P^\sigma(k)-k\,\Delta P^\sigma{}'(k)\right]\Big\}
\end{align}
(where $\Delta P^\sigma=P^R-P^L$), and compare with  \cite{Seto:2006hf}. First, $\delta(\vec{k}-\vec{k}')=\frac{\delta(k-k')}{k^2}\delta^{(2)}(\hat{n}-\hat{n}')$ where we  define
\begin{align}
\delta^{(2)}(\hat{n}-\hat{n}')=\frac{\delta(\theta-\theta')\,\delta(\phi-\phi')}{\sin\theta'}\,,
\end{align}
so that
\begin{align}
\int d^3 k\,\delta^{(3)}(\vec{k}-\vec{k}')=\int dk\sin\theta\,d\theta\,d\phi\,\delta(k-k')\delta^{(2)}(\hat{n}-\hat{n}')\,.
\end{align}
Then  we can write
\begin{eqnarray}
\langle h_R(k,\hat{n})h_R^*(k',\hat{n}')-h_L(k,\hat{n})h_L^*(k',\hat{n}')\rangle&=&\frac{\delta(k-k')\,\delta^{(2)}(\hat{n}-\hat{n}')}{8\pi\,k^5}\,\nonumber\\
&\times&\Big\{\Delta P^\sigma(k)-(\hat{n}\cdot \vec{v})\,\left[2\Delta P^\sigma(k)-k\,\Delta P^\sigma{}'(k)\right]\Big\}\nonumber\,.
\end{eqnarray}
We can compare with eq.~(\ref{ourcorrinV}) that can be written as
\begin{align}
&\langle h_R(f,\hat{n}){}h_R^*(f',\hat{n}') -h_L^*(f,\hat{n})h_L(f',\hat{n}') \rangle=\frac{\delta(f-f')}{f^4}\,\frac{\delta(\hat{n}-\hat{n}')}{\pi}V(f,\hat{n})\,\qquad f>0
\end{align}
so that
\begin{align}
V(f,\hat{n})=\frac{1}{8f}\Big\{\Delta P^\sigma(|f|)-(\hat{n}\cdot \vec{v})\,\left[2\Delta P^\sigma(|f|)-|f|\,\Delta P^\sigma{}'(|f|)\right]\Big\}\,.
\end{align}

Decomposing the quantity $V$ in spherical harmonics, by choosing the direction of $\vec{v}$ as the $z$ axis, we find
\begin{align}
&V_{00}=\frac{1}{8f}\Delta P^\sigma(|f|)\nonumber\,,\\
&V_{10}=-\frac{v}{8f}\left[2\Delta P^\sigma(|f|)-|f|\,\Delta P^\sigma{}'(|f|)\right]\,.
\end{align}
The work \cite{Seto:2006hf} defines the quantity $p$ that in our case, where we consider only the dipole contribution,   reads
\begin{align}
p=\frac{|V_{10}|}{I_t}\,.
\end{align}
Here $I_t$ is the total intensity, that in our regime is well approximated by  $|I_{00}|$, while \cite{Seto:2006hf}  defines $I$ as 
\begin{align}
&\frac{1}{2}\langle {}^{(S)}h_+(f,\hat{n})\,{}^{(S)}h^*_+(f',\hat{n}')+ {}^{(S)}h_\times(f,\hat{n})\,{}^{(S)}h^*_\times(f',\hat{n}')\rangle=\delta(f-f')\frac{\delta(\hat{n}-\hat{n}')}{4\pi}\,I(f)\nonumber\,, \\
&=\frac{f^4}{4}\langle h_R(f,\hat{n})h_R^*(f',\hat{n}')+h_L(f,\hat{n})h_L^*(f',\hat{n}')\rangle\simeq \frac{f^4}{4}\frac{\delta(k-k')\,\delta(\hat{n}-\hat{n}')}{8\pi\,k^5}\,\sum_\sigma P^\sigma(k)\,,
\end{align}
where we  only keep the monopole contribution from eq.~(\ref{finalhkhk}). So we get 
\begin{align}
I(f)=\frac{1}{8f}\sum_\sigma P^\sigma(|f|)\,,
\end{align}
and finally
\begin{align}
p\,=\,v\,\frac{|2\Delta P^\sigma(|f|)-|f|\,\Delta P^\sigma{}'(|f|)|}{\sum_\sigma P^\sigma(|f|)}\,.
\end{align}

Then using $P^\sigma\propto \Omega^\sigma/f^2$ with $\Omega_\sigma=\,$constant, we obtain
\begin{align}
p\,=\,4\,v\,\frac{|\sum_\lambda \lambda\Omega_\lambda|}{\sum_\lambda \Omega_\lambda}\,.
\end{align}
The final result of \cite{Seto:2006hf}, using $\sum_\lambda \Omega_\lambda=\Omega_{GW}$, can then be re-expressed as
\begin{align}
{}^{(S)}{\rm SNR}\,=\,4\times 10^{12} \, v\,  \left\vert \sum_\lambda \lambda\Omega_\lambda \right\vert \sqrt{\frac{T}{3\,{\rm years}}}\,.\label{comsnrse}
\end{align}

To compare with our findings, we  rewrite our result of eq.~\eqref{snr_final} as
\begin{align}
{\rm SNR}&\simeq\, 7.4\times10^{13}  \, v\,   \left\vert \sum_\lambda \lambda\Omega_\lambda\,h^2 \right\vert \sqrt{\frac{T}{3\,{\rm years}}}\simeq 3.6\times 10^{13}\, v\,    \left\vert \sum_\lambda \lambda\Omega_\lambda \right\vert \sqrt{\frac{T}{3\,{\rm years}}} \label{comsnrus}
\,,
\end{align}
 To conclude, our  result of eq. \eqref{comsnrus} is a factor of $10$ larger than the result of \cite{Seto:2006hf} in eq~\eqref{comsnrse}.

\section{Location and orientation of existing and forthcoming ground-based interferometers} 
\label{app:detectors}
\setcounter{equation}{0}

We take the Earth to be a perfect sphere of radius $R=6.371 \cdot 10^3 $km. We consider a Cartesian coordinate system with the origin located at the center of the Earth, and with the $z-$axis going in the direction of the North Pole. The $x-$axis goes in the direction of the point connecting the Earth Equator (latitude 0) and the Greenwich Meridian (longitude 0). The $y-$axis is then determined by ${\hat e}_y = {\hat e}_z \times {\hat e}_x$, and it is directed toward the point on the Equator at $90^\circ $E longitude. 

In Table \ref{tab:detectors} we provide the Cartesian coordinates (in the system that we have just defined) for a set of three unit vectors for each detector. The first unit vector is the position $\vec{x}_i$ of the $i-$th detector divided by $R$ (in practice, it is the unit-vector starting from the center of the Earth and pointing toward the center of the detector; by center we mean the point common to the two arms). The other two unit-vectors are the directions of each arm of the detector. Therefore, they are unit-vectors starting from the center of the interferometer, and lying on a plane that is tangent to the Earth at the point $\vec{x}_i$ (ignoring the curvature of the Earth on the scales of the interferometer arms). 

Ref. \cite{LigoGeo} provides the location (latitude and longitude) of the LIGO Hanford, the  LIGO Livingston, and the Virgo detectors, together with the direction that the arms of these detectors form with the North-South and East-West directions at that point on Earth. The same values for the KAGRA detector can be found  in Ref.~\cite{Akutsu:2017kpk}. The values in Table \ref{tab:detectors} are obtained from these data using basic trigonometry, and they are more convenient  for us, since the unit-vectors in the Table can be directly employed in our computations of Section~\ref{sec:ground_based}.

\begin{table}[h!]
\begin{center}
\begin{tabular}{|c|c|c|}
\hline
\multirow{3}{*}{LIGO Hanford} 
& Central location  & $ \left\{ -0.338 ,\; -0.600 ,\; 0.725 \right\} $  \\ 
\cline{2-3}
& First Arm  & $\left\{ -0.224 ,\; 0.799 ,\; 0.557 \right\}$   \\ 
\cline{2-3}
& Second Arm &  $\left\{ -0.914, 0.0261, -0.405 \right\}$ \\ 
\hline
\multirow{3}{*}{LIGO Livingston} 
& Central location  & $ \left\{ -0.0116 ,\; -0.861 ,\; 0.508 \right\}$ \\ 
\cline{2-3}
& First Arm  & $\left\{ -0.953 ,\; -0.144 ,\; -0.266 \right\}$  \\ 
\cline{2-3}
& Second Arm & $\left\{ 0.302 ,\; -0.488 ,\; -0.819 \right\}$ \\ 
\hline
\multirow{3}{*}{Virgo} 
& Central location  & $\left\{ 0.712 ,\; 0.132 ,\; 0.690 \right\}$ \\ 
\cline{2-3}
 & First Arm  & $\left\{ -0.701 ,\; 0.201 ,\; 0.684 \right\}$ \\ 
\cline{2-3}
 & Second Arm & $\left\{ -0.0485 ,\; -0.971 ,\; 0.236 \right\}$ \\ 
\hline
\multirow{3}{*}{KAGRA} 
& Central location  & $\left\{ -0.591 ,\; 0.546 ,\; 0.594 \right\}$ \\ 
\cline{2-3}
 & First Arm  & $\left\{ -0.390 ,\; -0.838 ,\; 0.382 \right\}$ \\ 
\cline{2-3}
 & Second Arm & $\left\{ 0.706 ,\; -0.00580 ,\; 0.709 \right\}$  \\ 
\hline
\end{tabular}
\caption{\it Cartesian coordinates of the unit-vectors specifying the positions of the interferometers and the direction of their arms, in the coordinate system described in this Appendix. For each detector, the distinction between the ``first'' and ``second'' arm is purely arbitrary, and plays no relevance in any computation. }
\label{tab:detectors}
\end{center}
\end{table}

\section{Analytics for  ground-based interferometer overlap functions }
\label{app:ground-analytic}
\setcounter{equation}{0}

In this Appendix we derive the analytic results (\ref{result-M-ground}) and (\ref{result-D-ground}) given in the main text. Let us start from the monopole response function (\ref{eq:response_ground}), that we rewrite as 
\begin{equation}
{\cal M}_{ij}^\lambda \left( k \right) =  D_i^{ab} \, D_j^{cd} \times \Gamma_{ab,cd,\lambda} \left( \kappa ,\, {\hat s}_{ij} \right) \;\;\;,\;\;\; \Gamma_{ab,cd,\lambda}^M \left( \kappa ,\, {\hat s} \right) \equiv  \int \frac{d \Omega_k}{4 \pi} \, 
{\rm e}^{i \kappa \,  {\hat k} \cdot {\hat s}} \, 
e_{ab, \lambda}( {\hat k})  e_{cd, \lambda}(- {\hat k}) \;. 
\label{GammaM}
\end{equation} 
The function $\Gamma$ must be a rank $4$ tensor, that is (separately) symmetric under the $a \leftrightarrow b$ and the  $c \leftrightarrow d$ interchange, as well as under $ab \leftrightarrow cd$. These symmetries enforce the structure 
\begin{eqnarray} 
\Gamma_{abcd,\lambda}^M \left( \kappa ,\, {\hat s} \right) &=& 
A_\lambda \left( \kappa \right) \delta_{ab} \delta_{cd} + B_\lambda \left( \kappa \right) \left( \delta_{ac} \delta_{bd} +  \delta_{ad} \delta_{bc} \right) 
\nonumber\\ 
&+& C_\lambda \left( \kappa \right) \left( \delta_{ab} {\hat s}_c {\hat s}_d + \delta_{cd} {\hat s}_a {\hat s}_b \right) \nonumber\\ 
&+& D_\lambda \left( \kappa  \right) \left( \delta_{ac} {\hat s}_b {\hat s}_d + \delta_{ad} {\hat s}_b {\hat s}_c + \delta_{bc} {\hat s}_a {\hat s}_d + \delta_{bd} {\hat s}_a {\hat s}_c \right) \nonumber\\ 
&+& E_\lambda \left( \kappa  \right) {\hat s}_a {\hat s}_b {\hat s}_c {\hat s}_d  
 \nonumber\\ 
 &+& F_\lambda \left( \kappa  \right)  \left(
 \delta_{ac} \epsilon_{bde} {\hat s}_e +
 \delta_{ad} \epsilon_{bce} {\hat s}_e +
 \delta_{bc} \epsilon_{ade} {\hat s}_e+ 
 \delta_{bd}  \epsilon_{ace} {\hat s}_e \right)
 \nonumber\\ 
 &+& 
G_\lambda \left( \kappa  \right)  \left(
 {\hat s}_{a} {\hat s}_{c} \epsilon_{bde} {\hat s}_e +
 {\hat s}_{a} {\hat s}_{d} \epsilon_{bce} {\hat s}_e +
 {\hat s}_{b} {\hat s}_{c} \epsilon_{ade} {\hat s}_e + 
 {\hat s}_{b} {\hat s}_{d}  \epsilon_{ace} {\hat s}_e 
 \right) \;, 
\label{Gamma-structure}
\end{eqnarray} 
where our goal is to find the scalar functions $A_\lambda \left( \kappa \right) ,\, \dots ,\, G_\lambda \left( \kappa  \right)$. To obtain these functions, we consider a set of independent contractions under which the angular integral of eq.~(\ref{GammaM}) becomes the integral of a scalar quantity, that can be immediately performed. Specifically, we perform the following contractions on the left hand side of eq.~(\ref{Gamma-structure}), 
\begin{eqnarray}
p_\lambda \left( \kappa \right)  &\equiv& \Gamma_{abcd,\lambda} \delta_{ab} \delta_{cd} = 0 \;, \nonumber\\ 
q_\lambda \left( \kappa \right)  &\equiv& \Gamma_{abcd,\lambda}  \left( \delta_{ac} \delta_{bd} +  \delta_{ad} \delta_{bc} \right)  =  2 j_0 \left( \kappa \right) \;, \nonumber\\ 
r_\lambda \left( \kappa \right) &\equiv& \Gamma_{abcd,\lambda}  \left( \delta_{ab} {\hat s}_c {\hat s}_d + \delta_{cd} {\hat s}_a {\hat s}_b \right) = 0 \;, \nonumber\\ 
{\hat s}_\lambda \left( \kappa \right) &\equiv& \Gamma_{abcd,\lambda}  \left( \delta_{ac} {\hat s}_b {\hat s}_d + \delta_{ad} {\hat s}_b {\hat s}_c + \delta_{bc} {\hat s}_a {\hat s}_d + \delta_{bd} {\hat s}_a {\hat s}_c \right)  =  \frac{4}{\kappa} \, j_1 \left( \kappa \right) \;, \nonumber\\ 
t_\lambda \left( \kappa \right) &\equiv& \Gamma_{abcd,\lambda}  {\hat s}_a {\hat s}_b {\hat s}_c {\hat s}_d =  \frac{2}{\kappa^2} \, j_2 \left( \kappa \right) \;, 
\nonumber\\ 
w_\lambda \left( \kappa \right) &\equiv& \Gamma_{abcd,\lambda}  \left( 
 \delta_{ac} \epsilon_{bde} {\hat s}_e +
 \delta_{ad} \epsilon_{bce} {\hat s}_e +
 \delta_{bc} \epsilon_{ade} {\hat s}_e+ 
 \delta_{bd}  \epsilon_{ace} {\hat s}_e 
 \right) = 4 \lambda j_1 \left( \kappa \right) \;, 
\nonumber\\ 
z_\lambda \left( \kappa \right) &\equiv& \Gamma_{abcd,\lambda}  \left(
 {\hat s}_{a} {\hat s}_{c} \epsilon_{bde} {\hat s}_e +
 {\hat s}_{a} {\hat s}_{d} \epsilon_{bce} {\hat s}_e +
 {\hat s}_{b} {\hat s}_{c} \epsilon_{ade} {\hat s}_e + 
 {\hat s}_{b} {\hat s}_{d}  \epsilon_{ace} {\hat s}_e 
\right) = \frac{4 \lambda}{\kappa} j_2 \left( \kappa \right) \;, 
\end{eqnarray} 
where we have also given the result of the integration in terms of spherical Bessel functions. Performing the same contractions on the right hand side of  eq.~(\ref{Gamma-structure}), and equating the results to the expressions that we have just found, we then obtain the system 
\begin{equation} 
\left( \begin{array}{ccccccc} 
9 & 6 & 6 & 4 & 1 & 0&0\\ 
6 & 24 & 4 & 16 & 2 & 0&0\\ 
6 & 4 & 8 & 8 & 2 & 0&0\\ 
4 & 16 & 8 & 24 & 4 & 0&0\\ 
1 & 2 & 2 & 4 & 1 & 0&0\\
0 & 0 & 0 & 0 & 0 & 40&8
\\
0 & 0 & 0 & 0 & 0 & 8&8
\end{array} \right) 
\left( \begin{array}{c} 
A_\lambda \\ B_\lambda \\ C_\lambda \\ D_\lambda \\ E_\lambda  \\ F_\lambda 
 \\ G_\lambda 
\end{array} \right) = 
\left( \begin{array}{c} 
p_\lambda \\ q_\lambda \\ r_\lambda \\ s_\lambda \\ t_\lambda 
\\w_\lambda  \\z_\lambda 
\end{array} \right) \;. 
\end{equation} 

This linear system is solved by 
\begin{eqnarray}
&& A_\lambda = - \frac{j_1 \left( \kappa \right)}{4 \kappa} + \frac{1+ \kappa^2}{4 \kappa^2} \, j_2 \left( \kappa \right) \;\;,\;\; 
B_\lambda =  \frac{j_1 \left( \kappa \right)}{4 \kappa} + \frac{1-\kappa^2}{4 \kappa^2} \, j_2 \left( \kappa \right) \;, 
\nonumber\\ 
&& C_\lambda =  \frac{j_1 \left( \kappa \right)}{4 \kappa} - \frac{5+\kappa^2}{4 \kappa^2} \, j_2 \left( \kappa \right) \;\;,\;\; 
D_\lambda = \frac{j_1 \left( \kappa \right)}{4 \kappa}-\frac{5- \kappa^2}{4 \kappa^2} \, j_2 \left( \kappa \right)  \;, 
\nonumber\\ 
&& E_\lambda =  \frac{-7 j_1 \left( \kappa \right)}{4 \kappa}+\frac{35- \kappa^2}{4 \kappa^2} \, j_2 \left( \kappa \right)  \;, 
\nonumber\\ 
&& F_\lambda = \lambda \left[  \frac{j_1 \left( \kappa \right)}{8} -  \frac{j_2 \left( \kappa \right)}{8 \kappa} \right] \;\;,\;\;
G_\lambda = \lambda \left[  - \frac{j_1 \left( \kappa \right)}{8} + 5 \frac{j_2 \left( \kappa \right)}{8 \kappa} \right] \;. 
\label{A-G-sol}
\end{eqnarray} 

We have thus fully obtain the analytic expression for (\ref{Gamma-structure}). Once we contract with the detector functions $D_i^{ab}D_j^{ab}$, the terms proportional to $A \left( \kappa \right)$ and $C \left( \kappa \right)$ do not contribute to the response function (\ref{GammaM}) due to the fact that these operators are traceless. The remaining terms give rise to the expression (\ref{result-M-ground}), upon the relabelling $2 B_\lambda \to f_A ,\; 4 D_\lambda \to f_B ,\;  E_\lambda \to f_C ,\; 4 F_\lambda \to \lambda \, f_D ,\;  4 G_\lambda \to \lambda \, f_E $.

Let us now move to the dipole response function  (\ref{eq:response_ground}), that we rewrite as 
\begin{equation}
{\cal D}_{ij}^\lambda \left( k \right) =  D_i^{ab} \, D_j^{cd} \times \Gamma_{ab,cd,\lambda}^D \left( \kappa ,\, {\hat s}_{ij} \right) \;\;\;,\;\;\; \Gamma_{ab,cd,\lambda}^D \left( \kappa ,\, {\hat s} ,\, {\hat v} \right) \equiv i {\hat v} \cdot \int \frac{d \Omega_k}{4 \pi} \, 
{\rm e}^{i \kappa \,  {\hat k} \cdot {\hat s}} \, 
e_{ab, \lambda}( {\hat k})  e_{cd, \lambda}(- {\hat k}) \;. 
\label{GammaD}
\end{equation} 
A direct comparison between eqs. (\ref{GammaM}) and (\ref{GammaD}) shows that the function $\Gamma_D$ can be obtained from a derivative of the function $\Gamma^M$ that we just computed 
\begin{eqnarray}
\Gamma_{ab,cd,\lambda}^D  \left( \kappa ,\, {\hat s} ,\, {\hat v} \right)  
= \left[ \frac{1}{\kappa} \,  {\hat v}_i \, \frac{\partial}{\partial s_i} {\tilde \Gamma}_{ab,cd,\lambda}^M \left( \kappa ,\, \vec{s} \right) 
\right] \Big\vert_{s=1} 
\label{GammaM-GammaD}
\end{eqnarray} 

Before differentiating, we need to promote the expression in eq.~(\ref{Gamma-structure}) to be a function of a vector $\vec{s}$ of arbitrary magnitude. This can be immediately done from the result we obtained by noting (from the definition in eq.~(\ref{GammaM})) that the magnitude can be absorbed in $\kappa$. Therefore, for example, the term proportional to $C$ becomes 
\begin{equation}  
{\tilde \Gamma}_{ab,cd,\lambda}^M \left( \kappa ,\, \vec{s} \right) = \dots +  C_\lambda \left( \kappa \, s \right) \left( \frac{\delta_{ab}  \, \vec{s}_c \, \vec{s}_d + \delta_{cd} \, \vec{s}_a \, \vec{s}_b}{s^2} \right) + \dots \;. 
\end{equation} 
Taking this into account, inserting the result (\ref{Gamma-structure}) - (\ref{A-G-sol}) in eq.~(\ref{GammaM-GammaD}) leads to the analytic expression for $\Gamma_{ab,cd,\lambda}^D  \left( \kappa ,\, {\hat s} ,\, {\hat v} \right)$. This expression, once contracted with $D_i^{ab} \, D_j^{cd}$ leads to the result (\ref{result-D-ground}).

\end{appendix}

\bigskip

\noindent

\addcontentsline{toc}{section}{References}
\bibliographystyle{utphys}

\bibliography{refsdipole}

\end{document}